\documentclass[default, twocolumn]{aastex7}
\usepackage{amsmath}
\usepackage{siunitx}
\usepackage{soul}

\newcommand{\chicagoastro}{Department of Astronomy and Astrophysics, University of Chicago, Chicago, IL, 60637, USA}
\newcommand{\chicagophysics}{Department of Physics, University of Chicago, Chicago, IL 60637, USA}
\newcommand{\princetonphysics}{Joseph Henry Laboratories of Physics, Jadwin Hall, Princeton University, Princeton, NJ 08544, USA}
\newcommand{\ucsdastro}{Department Astronomy and Astrophysics, University of California San Diego, La Jolla, CA, 92093, USA}
\newcommand{\ucsdphysics}{Department of Physics, University of California San Diego, La Jolla, CA, 92093 USA}
\newcommand{\cornellphysics}{Department of Physics, Cornell University, Ithaca, NY, 14853, USA}
\newcommand{\catolicafisica}{Instituto de Astrof\'isica and Centro de Astro-Ingenier\'ia, Facultad de F\'isica, Pontificia Universidad Cat\'olica de Chile, Santiago, Chile}
\newcommand{\upennphysicsastro}{Department of Physics and Astronomy, University of Pennsylvania, Philadelphia, PA 19104, USA}
\newcommand{\kipacstanford}{Kavli Institute for Particle Astrophysics and Cosmology, Stanford, CA 94305, USA}
\newcommand{\ucriverside}{Center for Experimental Cosmology and Instrumentation, Department of Physics and Astronomy, University of California, Riverside, CA 92521, USA}

\begin{document}

\title{The Simons Observatory: Commissioning of the Mid-Frequency Small Aperture Telescopes}

\author[0000-0003-1248-9563]{Kathleen Harrington}
\affiliation{High Energy Physics Division, Argonne National Laboratory, Lemont, IL 60439, USA}
\affiliation{\chicagoastro}
\email[show]{kharrington@anl.gov}  

\author[0009-0009-0876-9168]{Remington Gerras}
\affiliation{Department of Physics and Astronomy, University of Southern California, Los Angeles, CA
90089-1483 USA}
\email{gerras@usc.edu}  

\author[0000-0001-7225-6679]{Nicholas Galitzki}
\affiliation{Department of Physics, University of Texas at Austin, Austin, TX 78712, USA}
\affiliation{Weinberg Institute for Theoretical Physics, Texas Center for Cosmology and Astroparticle Physics, Austin, TX 78712, USA}
\email{nicholas.galitzki@austin.utexas.edu} 

\author[0009-0003-5814-2087]{Samuel Day-Weiss}
\affiliation{\princetonphysics}
\email{dayweiss@princeton.edu} 

\author[0000-0002-3644-2009]{Yoshinori Sueno}
\affiliation{\princetonphysics}
\affiliation{Kavli IPMU (WPI), UTIAS, The University of Tokyo, Kashiwa, Chiba 277-8583, Japan}
\email{yoshinorisueno@gmail.com} 

\author[0000-0003-1942-1334]{Thomas Alford}
\affiliation{\chicagophysics}
\email{tdalford@uchicago.edu} 

\author[0009-0009-9806-2317]{Michael J. Randall}
\affiliation{Department of Physics, University of California - Berkeley, Berkeley, CA 94720, USA}
\email{mjrandall@berkeley.edu} 

\author[0000-0003-0221-2130]{Kyohei Yamada}
\affiliation{\princetonphysics}
\email{ykyohei7@gmail.com} 

\author[0000-0001-7480-4341]{Maximiliano Silva-Feaver}
\affiliation{Department of Physics, Yale University, New Haven, CT 06511 USA}
\email{maximiliano.silva-feaver@yale.edu} 

\author[0000-0001-5068-1295]{Kevin Crowley}
\affiliation{Department of Astronomy and Astrophysics, University of California San Diego, La Jolla, CA 92093}
\email{ktcrowley@ucsd.edu} 

\author[0000-0002-0400-7555]{Shunsuke Adachi}
\affiliation{Okayama University, Department of Physics, Okayama 700-8530, Japan}
\affiliation{Kavli Institute for the Physics and Mathematics of the Universe (WPI), UTIAS, The University of Tokyo, Kashiwa, Chiba, 277-8583, Japan}
\affiliation{High Energy Accelerator Research Organization (KEK), Tsukuba, 305-0801, Japan}
\email{shadachi@s.okayama-u.ac.jp}

\author[0000-0002-5736-5524]{Alexandre E Adler}
\affiliation{Department of Physics, University of California Berkeley, Berkeley, CA 94720, USA}
\affiliation{Computational Cosmology Center, Lawrence Berkeley National Laboratory, Berkeley, CA 94720, USA}
\affiliation{CNRS-UCB International Research Laboratory, Centre Pierre Binetruy, IRL 2007, CPB- IN2P3, Berkeley, US}
\email{aadler@lbl.gov}

\author[0009-0006-8601-6696]{Prakamya Agrawal}
\affiliation{\ucsdastro}
\email{p9agrawal@ucsd.edu}

\author[0000-0002-1035-1854]{Simone Aiola}
\affiliation{Center for Computational Astrophysics, Flatiron Institute, New York, NY, 10010, USA}
\email{saiola@flatironinstitute.org}

\author[0000-0002-4598-9719]{David Alonso}
\affiliation{Department of Physics, University of Oxford, Denys Wilkinson Building, Keble Road, Oxford OX1 3RH, United Kingdom}
\email{david.alonso@physics.ox.ac.uk}

\author[0000-0002-3407-5305]{Kam Arnold}
\affiliation{\ucsdastro}
\affiliation{\ucsdphysics}
\email{arnold@ucsd.edu}

\author[0000-0002-8132-4896]{Susanna Azzoni}
\affiliation{\princetonphysics}
\email{sazzoni@princeton.edu}

\author[0000-0002-7888-6222]{Andrew Bazarko}
\affiliation{\princetonphysics}
\email{bazarko@princeton.edu}

\author[0000-0002-9763-1663]{Sanah Bhimani}
\affiliation{\chicagoastro}
\email{sanahbhimani@gmail.com}

\author[0000-0002-1493-2963]{Simon Biquard}
\affiliation{Jodrell Bank Centre for Astrophysics, Department of Physics and Astronomy, University of Manchester, Manchester M13 9PL, UK}
\email{simon.biquard@manchester.ac.uk}

\author[0009-0008-4312-6814]{Bryce Bixler}
\affiliation{\ucsdphysics}
\email{bbixler@ucsd.edu}

\author[0000-0002-1327-1921]{Josh Borrow}
\affiliation{\upennphysicsastro}
\email{josh@joshborrow.com}

\author[0000-0002-0370-8077]{Michael L. Brown}
\affiliation{Jodrell Bank Centre for Astrophysics, Department of Physics and Astronomy, University of Manchester, Manchester M13 9PL, UK}
\email{m.l.brown@manchester.ac.uk}

\author[0000-0003-0837-0068]{Erminia Calabrese}
\affiliation{Department of Physics and Astronomy, School of Physical, Chemical and Environmental Sciences, Cardiff University, The Parade, Cardiff, Wales, UK CF24 3AA}
\email{calabresee@cardiff.ac.uk}

\author[0000-0002-3266-857X]{Yuji Chinone}
\affiliation{International Center for Quantum-field Measurement Systems for Studies of the Universe and Particles (WPI-QUP), High Energy Accelerator Research Organization (KEK), 1-1 Oho, Tsukuba, Ibaraki 305-0801, Japan}
\email{chinoney@gmail.com}

\author[0000-0002-9113-7058]{Steve K. Choi}
\affiliation{\ucriverside}
\email{schoi@ucr.edu}

\author[0009-0006-7382-1434]{Nadia Dachlythra}
\affiliation{Department of Physics, University of Milano-Bicocca, 20126, Milano, Italy}
\email{konstantina.dachlythra@unimib.it}

\author[0000-0002-3169-9761]{Mark Devlin}
\affiliation{\upennphysicsastro}
\email{devlin@physics.upenn.edu}

\author[0000-0002-1940-4289]{Simon Dicker}
\affiliation{\upennphysicsastro}
\email{sdicker@physics.upenn.edu}

\author[0009-0006-8427-6259]{Peter N. Dow}
\affiliation{Department of Astronomy, University of Virginia, Charlottesville, VA 22904, USA}
\email{pd3cx@virginia.edu}

\author[0000-0002-9693-4478]{Shannon M. Duff}
\affiliation{Quantum Sensors Division, National Institute of Standards and Technology, Boulder, CO 80305, USA}
\email{shannon.duff@nist.gov}

\author[0000-0002-7450-2586]{Jo Dunkley}
\affiliation{\princetonphysics}
\affiliation{Department of Astrophysical Sciences, Peyton Hall, Princeton University, Princeton, NJ 08544, USA}
\email{jdunkley@princeton.edu}

\author[0000-0003-3892-1860]{Rolando Dunner}
\affiliation{\catolicafisica}
\email{rdunnerp@uc.cl}

\author[0000-0002-9962-2058]{Daniel Dutcher}
\affiliation{\princetonphysics}
\email{ddutcher@uchicago.edu}

\author[0000-0001-5471-3434]{Hamza El Bouhargani}
\affiliation{\princetonphysics}
\email{helbouha@princeton.edu}

\author[0000-0002-1419-0031]{Josquin Errard}
\affiliation{Universit\'{e} Paris Cit\'{e}, CNRS, Astroparticule et Cosmologie, F-75013 Paris, France}
\email{josquin@apc.in2p3.fr}

\author[0000-0002-7145-1824]{Allen Foster}
\affiliation{\princetonphysics}
\email{amfoster@princeton.edu}

\author[0000-0001-8159-8208]{Ken Ganga}
\affiliation{Universit\'{e} Paris Cit\'{e}, CNRS, Astroparticule et Cosmologie, F-75013 Paris, France}
\email{ken.ganga@u-paris.fr}

\author[0000-0001-9880-3634]{John Groh}
\affiliation{Physics Division, Lawrence Berkeley National Laboratory, Berkeley, CA 94720, USA}
\email{john.groh@lbl.gov}

\author[0000-0002-1697-3080]{Yilun Guan}
\affiliation{Dunlap Institute for Astronomy and Astrophysics, University of Toronto, M5S 3H4, Canada}
\email{yilun.guan@utoronto.ca}

\author[0000-0003-1760-0355]{Jon E.~Gudmundsson}
\affiliation{Science Institute, University of Iceland, 107 Reykjavik, Iceland}
\affiliation{The Oskar Klein Centre for Cosmoparticle Physics, Department of Physics, Stockholm University, AlbaNova, SE-106 91 Stockholm, Sweden}
\email{jegudmunds@gmail.com}

\author[0000-0001-6519-502X]{Saianeesh Haridas}
\affiliation{\upennphysicsastro}
\email{haridas@sas.upenn.edu}

\author[0000-0003-1443-1082]{Masaya Hasegawa}
\affiliation{Institute of Particle and Nuclear Studies (IPNS), High Energy Accelerator Research Organization (KEK), 1-1 Oho,Tsukuba, Ibaraki 305-0801, Japan}
\affiliation{International Center for Quantum-field Measurement Systems for Studies of the Universe and Particles (WPI-QUP), High Energy Accelerator Research Organization (KEK), 1-1 Oho, Tsukuba, Ibaraki 305-0801, Japan}
\affiliation{Particle and Nuclear Physics Program, The Graduate University for Advanced Studies (SOKENDAI), Shonan Village, Hayama, Kanagawa 240-0193, Japan}
\email{masaya.hasegawa@kek.jp}

\author[0000-0002-2408-9201]{Matthew Hasselfield}
\affiliation{Center for Computational Astrophysics, Flatiron Institute, New York, NY, 10010, USA}
\email{mhasselfield@flatironinstitute.org}

\author[0000-0001-7878-4229]{Shawn W. Henderson}
\affiliation{\kipacstanford}
\affiliation{SLAC National Accelerator Laboratory, Menlo Park, CA 94025, USA}
\email{shawn@slac.stanford.edu}

\author[0000-0002-4765-3426]{Carlos Herv\'ias-Caimapo}
\affiliation{\catolicafisica }
\email{carlos.hervias@uc.cl}

\author[0000-0003-1690-6678]{Adam D.~Hincks}
\affiliation{David A. Dunlap Department of Astronomy \& Astrophysics, University of Toronto, Toronto, ON M5S 3H4, Canada}
\affiliation{Specola Vaticana (Vatican Observatory), V-00120 Vatican City, Vatican City State}
\email{adam.hincks@utoronto.ca}

\author[0000-0002-0965-7864]{Ren\'ee Hlo\v{z}ek}
\affiliation{Dunlap Institute for Astronomy and Astrophysics, University of Toronto, M5S 3H4, Canada}
\affiliation{David A. Dunlap Department of Astronomy \& Astrophysics, University of Toronto, Toronto, ON M5S 3H4, Canada}
\email{hlozek@dunlap.utoronto.ca}

\author[0000-0002-2781-9302]{Johannes Hubmayr}
\affiliation{Quantum Sensors Division, National Institute of Standards and Technology, Boulder, CO 80305, USA}
\email{johanneshubmayr@gmail.com}

\author[0000-0002-6898-8938]{Bradley R. Johnson}
\affiliation{Department of Astronomy, University of Virginia, Charlottesville, VA 22904, USA}
\email{bradley.johnson@virginia.edu}

\author[0009-0001-3477-5141]{Yutaro Kasai}
\affiliation{Department of Physics, Faculty of Science, Kyoto University, Kyoto 606-8502, Japan}
\email{kasai.yuutarou.57a@st.kyoto-u.ac.jp}

\author[0000-0003-3118-5514]{Brian Keating}
\affiliation{\ucsdphysics}
\email{bkeating@ucsd.edu}

\author[0000-0002-2978-7957]{Ben Keller}
\affiliation{\cornellphysics}
\email{bdk54@cornell.edu}

\author[0000-0003-3510-7134]{Theodore Kisner}
\affiliation{Computational Cosmology Center, Lawrence Berkeley National Laboratory, Berkeley, CA 94720, USA}
\affiliation{Space Sciences Lab, University of California Berkeley, Berkeley, CA 94720, USA}
\email{tskisner@lbl.gov}

\author[0000-0003-0744-2808]{Brian J. Koopman}
\affiliation{Wright Laboratory, Department of Physics, Yale University, New Haven, CT 06511, USA}
\email{brian.koopman@yale.edu}

\author[0009-0004-9631-2451]{Akito Kusaka}
\affiliation{Department of Physics, The University of Tokyo, Tokyo 113-0033, Japan}
\affiliation{Research Center for the Early Universe, School of Science, The University of Tokyo, Tokyo 113-0033, Japan}
\affiliation{Kavli Institute for the Physics and Mathematics of the Universe (WPI), UTIAS, The University of Tokyo, Kashiwa, Chiba, 277-8583, Japan}
\affiliation{Physics Division, Lawrence Berkeley National Laboratory, Berkeley, CA 94720, USA}
\email{akusaka@phys.s.u-tokyo.ac.jp}

\author[0000-0003-3106-3218]{Adrian Lee}
\affiliation{Department of Physics, University of California Berkeley, Berkeley, CA 94720, USA}
\affiliation{Physics Division, Lawrence Berkeley National Laboratory, Berkeley, CA 94720, USA}
\email{adrian.lee@berkeley.edu}

\author[0000-0003-1581-1626]{JB Lloyd}
\affiliation{Department of Physics, University of Texas at Austin, Austin, TX 78712, USA}
\affiliation{Weinberg Institute for Theoretical Physics, Texas Center for Cosmology and Astroparticle Physics, Austin, TX 78712, USA}
\email{jlloydiii@utexas.edu}

\author[0000-0003-1200-9179]{Anto Lonappan}
\affiliation{\ucsdphysics}
\email{alonappan@ucsd.edu}

\author[0000-0002-3800-5558]{Marius Lungu}
\affiliation{The Simons Observatory, New York, NY, 10001, USA}
\email{mvlungu@gmail.com}

\author[0009-0000-1028-3524]{Aashrita Mangu}
\affiliation{\chicagophysics}
\email{amangu@uchicago.edu}

\author[0009-0008-5935-5742]{Michael McCrackan}
\affiliation{Wright Laboratory, Department of Physics, Yale University, New Haven, CT 06511, USA}
\email{michael.mccrackan@yale.edu}

\author[0000-0002-7245-4541]{Jeff McMahon}
\affiliation{\chicagoastro}
\affiliation{\chicagophysics}
\affiliation{Enrico Fermi Institute, University of Chicago, Chicago, IL, 60637, USA}
\affiliation{Kavli Institute for Cosmological Physics, University of Chicago, Chicago, IL, 60637, USA}
\email{jjm@uchicago.edu}

\author[0000-0002-7340-9291]{Jenna Moore}
\affiliation{Department of Physics, Duke University, Durham, NC 27708, USA}
\email{jenna.moore@duke.edu}

\author[0000-0002-5564-997X]{Thomas W. Morris}
\affiliation{Wright Laboratory, Department of Physics, Yale University, New Haven, CT 06511, USA}
\email{thomas.w.morris@yale.edu}

\author[0000-0002-6300-1495]{Hironobu Nakata}
\affiliation{Department of Physics, Faculty of Science, Kyoto University, Kyoto 606-8502, Japan}
\email{zhongtianjiaxin@gmail.com}

\author[0000-0002-8307-5088]{Federico Nati}
\affiliation{Department of Physics, University of Milano-Bicocca, 20126, Milano, Italy}
\email{federico.nati@unimib.it}

\author[0000-0002-7333-5552]{Laura Newburgh}
\affiliation{Wright Laboratory, Department of Physics, Yale University, New Haven, CT 06511, USA}
\email{laura.newburgh@yale.edu}

\author[0000-0002-7575-8145]{David V.~Nguyen}
\affiliation{Wright Laboratory, Department of Physics, Yale University, New Haven, CT 06511, USA}
\email{david.nguyen@yale.edu}

\author[0000-0003-1842-8104]{John Orlowski-Scherer}
\affiliation{\upennphysicsastro}
\email{jorlo@sas.upenn.edu}

\author[0000-0002-9828-3525]{Lyman Page}
\affiliation{\princetonphysics}
\email{page@princeton.edu}

\author[0009-0000-6712-1307]{Ioannis Paraskevakos}
\affiliation{Research Computing, Princeton University, Princeton, NJ, 08544, USA}
\email{iparask@princeton.edu}

\author[0000-0002-9516-3245]{Tristan Pinsonneault-Marotte}
\affiliation{\kipacstanford}
\affiliation{SLAC National Accelerator Laboratory, Menlo Park, CA 94025, USA}
\email{tristpm@slac.stanford.edu}

\author[0000-0003-3484-5645]{Erik Rosenberg}
\affiliation{Jodrell Bank Centre for Astrophysics, Department of Physics and Astronomy, University of Manchester, Manchester M13 9PL, UK}
\email{erik.rosenberg@manchester.ac.uk}

\author[0000-0001-6389-0117]{Yuki Sakurai}
\affiliation{Suwa University of Science, Department of Mechanical and Electrical Engineering, Chino, Nagano, 291-0292, Japan}
\affiliation{Kavli Institute for the Physics and Mathematics of the Universe (WPI), UTIAS, The University of Tokyo, Kashiwa, Chiba, 277-8583, Japan}
\email{sakurai_yuki@rs.sus.ac.jp}

\author[0009-0001-6039-7834]{Goureesankar Sathyanathan}
\affiliation{\ucriverside}
\email{goureesankar.sathyanathan@email.ucr.edu}

\author[0000-0002-6452-4220]{Thomas P. Satterthwaite}
\affiliation{Department of Physics, Stanford University, Stanford, CA 94305, USA}
\affiliation{\kipacstanford}
\email{tpsatt@stanford.edu}

\author[0000-0001-5680-4989]{Yudai Seino}
\affiliation{\princetonphysics}
\email{ys9136@princeton.edu}

\author[0000-0001-5644-8750]{Elle Shaw}
\affiliation{Department of Physics, University of Texas at Austin, Austin, TX 78712, USA}
\affiliation{Weinberg Institute for Theoretical Physics, Texas Center for Cosmology and Astroparticle Physics, Austin, TX 78712, USA}
\email{elle.shaw@austin.utexas.edu}

\author[0009-0000-0668-3584]{Sara M. Simon}
\affiliation{Cosmic Frontier Division, Fermi National Accelerator Laboratory, Batavia, IL, 60510, USA}
\affiliation{\chicagoastro}
\email{simon.sara.m@gmail.com}

\author[0000-0002-7020-7301]{Suzanne T. Staggs}
\affiliation{\princetonphysics}
\email{staggs@princeton.edu}

\author[0009-0007-7435-9082]{Junna Sugiyama}
\affiliation{\ucriverside}
\email{junnas@ucr.edu}

\author[0000-0001-6816-8123]{Junya Suzuki}
\affiliation{Department of Physics, Faculty of Science, Kyoto University, Kyoto 606-8502, Japan}
\email{suzuki.junya.4r@kyoto-u.ac.jp}

\author[0000-0001-9461-7519]{Satoru Takakura}
\affiliation{Department of Physics, The University of Tokyo, Tokyo 113-0033, Japan}
\email{satoru.takakura@phys.s.u-tokyo.ac.jp}

\author[0000-0003-2439-2611]{Osamu Tajima}
\affiliation{Department of Physics, Faculty of Science, Kyoto University, Kyoto 606-8502, Japan}
\affiliation{High Energy Accelerator Research Organization (KEK), Tsukuba, 305-0801, Japan}
\affiliation{Kavli Institute for the Physics and Mathematics of the Universe (WPI), UTIAS, The University of Tokyo, Kashiwa, Chiba, 277-8583, Japan}
\email{tajima.osamu.8a@kyoto-u.ac.jp}

\author[0000-0001-9528-8147]{Alex Thomas}
\affiliation{\chicagoastro}
\affiliation{Kavli Institute for Cosmological Physics, University of Chicago, Chicago, IL, 60637, USA}
\email{agthomas@uchicago.edu}

\author[0000-0003-2244-9530]{Daniel B. Thomas}
\affiliation{Jodrell Bank Centre for Astrophysics, Department of Physics and Astronomy, University of Manchester, Manchester M13 9PL, UK}
\email{dan.b.thomas1@gmail.com}

\author[0000-0002-1667-2544]{Tran Tsan}
\affiliation{Physics Division, Lawrence Berkeley National Laboratory, Berkeley, CA 94720, USA}
\email{ttsan@lbl.gov}

\author[0000-0002-2105-7589]{Eve Vavagiakis}
\affiliation{Department of Physics, Duke University, Durham, NC 27708, USA}
\affiliation{\cornellphysics}
\email{eve.vavagiakis@duke.edu}

\author[0000-0002-8710-0914]{Yuhan Wang}
\affiliation{\cornellphysics}
\email{yuhanwyhan@gmail.com}

\author[0000-0001-7828-7257]{John Wilson}
\affiliation{Department of Astronomy, University of Virginia, Charlottesville, VA 22904, USA}
\email{jcw6z@virginia.edu}
\collaboration{all}{The Simons Observatory Collaboration}

\newcommand{\etafninety}{$\eta_{\mathrm{opt}} \in [0.17, 0.31]$ }
\newcommand{\etafonefifty}{$\eta_{\mathrm{opt}} \in [0.28, 0.48]$ }
\newcommand{\loadingSATone}{2.9--4.2\,pW (4.1--6.2\,pW) }
\newcommand{\loadingSATthree}{2.2--3.0\,pW (3.2--5.2\,pW) }

\begin{abstract}
The Simons Observatory (SO) is a cosmic microwave background (CMB) survey experiment observing from an altitude of 5,200\;m in the Atacama Desert in Chile. 
The SO Small Aperture Telescopes (SATs) will cover six spectral bands between 30 and 280\;GHz primarily to search for evidence of primordial gravitational waves in the large angular scale $B$-modes of the CMB polarization. 
The first two Mid-Frequency (MF) SATs, operating in the 90 and 150\,GHz frequency bands, began commissioning observations in late 2023, and we present the results of those observations here. 
Each MF SAT operates with approximately 9,400 biased transition-edge sensors (TES) at 1\,mm of precipitable water vapor (PWV), representing a typical 80\% yield.
Planet observations confirm the telescope optics performance, calibrate the instrument response, and measure the end-to-end optical efficiency of the SATs. 
The realized instrument efficiencies are within the range expected based on instrument model predictions. 
In nominal observing conditions, the noise equivalent temperatures from both telescopes combined are $2.4~\si{\micro\kelvin}\mathrm{_{CMB}\sqrt{s}}$ at 90~GHz and $3.2~\si{\micro\kelvin}\mathrm{_{CMB}\sqrt{s}}$ at 150~GHz, substantially outperforming the baseline instantaneous sensitivity targets for the observatory.

\end{abstract}
 

\section{Introduction \label{sec:intro}} 
\setcounter{footnote}{0}

Understanding the origin and evolution of the Universe is one of the challenges of modern physics. Cosmic inflation is the leading theoretical framework for describing the dynamics of the early Universe~\citep[e.g.,][]{snowmass_inflation_2022}; these dynamics explain the flatness of spacetime, the horizon problem, and the near-scale-invariant spectrum of primordial density perturbations~\citep{guth:1981,Mukhanov/Chibisov:1981,starobinsky:1982, Linde1982b, Bardeen1983}. A key prediction of inflation is a background of primordial gravitational waves \citep{starobinsky:1979}, that imprint a characteristic parity-odd ($B$-mode) pattern in the polarization of the cosmic microwave background (CMB) at degree angular scales, with an amplitude parameterized by the tensor-to-scalar ratio~$r$~\citep[e.g.,][]{polnarev:1985, Crittenden1993, kami97,zald97}. Detection of this signal would directly probe physics at energy scales of order $10^{16}$~GeV, well beyond any terrestrial accelerator and inaccessible by any other known observational means. (For additional discussion on inflation see, for example, \citealt{Weinberg2008Cosmology, Baumann2022Cosmology} and references therein).

Recent searches for the $B$-mode signal sourced from primordial gravitational waves have been conducted by low-resolution telescopes such as BICEP~\citep{BK_2021}, SPIDER~\citep{spider_2021}, CLASS~\citep{Li_2025}, ABS~\citep{Kusaka_2018} and the \textit{Planck} satellite~\citep{planck_inflation}; as well as higher-resolution instruments such as \textsc{Polarbear}~\citep{Adachi_2022} and the South Pole Telescope~\citep{spt_bmodes_2025}. The most stringent constraint on the tensor-to-scalar ratio from the $B$-mode signal is $r_{0.05}<0.036$~(95\% C.L.)\footnote{The subscript indicates that the ratio is evaluated at the co-moving wavenumber  $k=0.05$~Mpc$^{-1}$.} from \cite{BK_2021} (see also \citealt{Tristram2022}).
 
To search for $B$-modes and pursue additional science goals, the Simons Observatory (SO) has built a suite of state-of-the-art CMB telescopes located in the Atacama Desert of Northern Chile, near the summit of Cerro Toco, at an elevation of 5,200 meters, an ideal ground-based location to study the millimeter signal from the CMB~\citep{site_Bustos_2014}. The observatory comprises one 6~meter Large Aperture Telescope (LAT)~\citep{Zhu_2021,ASO_LAT_2025} and multiple Small Aperture Telescopes (SATs)~\citep{Galitzki_2018, Galitzki2024}, designed to probe fundamental questions about the early universe, including inflation, neutrino physics, and the nature of dark matter and dark energy as well as to advance millimeter-wave astrophysics through studies of galaxy clusters via the Sunyaev-Zel'dovich effect, the time-domain millimeter sky, the diffuse Galactic interstellar medium, and the population and properties of extragalactic sources~(\citealt{SO2019_Forecasting}, \citealt{galactic_forecasting_2022, ASO_LAT_2025, sosats_extended2025}). 

The SAT is a refracting telescope; its design includes a 42~cm aperture stop, a $35^\circ$~field of view, a continuously rotating cryogenic half-wave plate~(HWP, \citealt{Yamada2026}), and several layers of optical shielding, all to optimize the telescopes for measurements of the degree-angular-scale polarization of the CMB~\citep{Galitzki2024}. The initial configuration of SO includes three SATs operating with dichroic polarization-sensitive detectors:  two Mid-Frequency (MF) telescopes, with frequency passbands centered near 90 and 150~GHz, and one Ultra-High-Frequency (UHF) telescope, with passbands centered near 220 and 280~GHz. The observing frequencies lie within windows of high atmospheric transmission and were chosen to maximize sensitivity to the CMB while also enabling foreground component separation. The passbands will be denoted by f090, f150, f220, and f280 throughout the rest of this paper.  Each of these initial SATs includes over 12,000 microfabricated transition-edge sensor (TES) detectors~\citep{Duff_2024, Dutcher_2024}. In the future, the SO will deploy three more SATs: two to further improve sensitivity in the MF range and an additional Low-Frequency (LF) telescope with observing frequencies centered near 30 and 40~GHz~\citep{sosats_extended2025}. The MF SATs began commissioning observations in late 2023. The UHF SAT began commissioning observations in late 2025 and will be discussed in future work.
This paper presents the commissioning plan and results of the first two MF SATs, including system integration, calibration procedures, initial performance metrics, and lessons learned. 

\section{Commissioning Requirements and Planning}
\label{sec:requirements}

To make maps of the CMB and its polarization, the SO SATs must be able to survey the sky with a dedicated ``CMB scan strategy,” maintaining their unprecedented sensitivity year after year. The commissioning goals for each telescope are  (1) to demonstrate that the integrated telescope meets or surpasses the instantaneous white noise performance assumed in \citeauthor{SO2019_Forecasting} (\citeyear{SO2019_Forecasting}; SO2019 hereafter), (2) to optimize the scan parameters for the CMB scan strategy and (3) to demonstrate the capability to maintain sustained operations that meet or exceed target observing efficiencies.  

Completion of the commissioning of the first two MF SATs marked a critical milestone in the SO program;   this paper presents the associated observations and analyses.   Future work will describe the results of the next step in evaluating the SATs:   analysis of  data  from  $\sim~4$ months of   ``initial science observations'' (ISO), with the goal of demonstrating that the CMB maps are consistent with the performance described in SO2019.


\begin{deluxetable*}{lcccccc}
\tablecaption{Instantaneous array noise (noise equivalent temperatures, NETs) requirements (top two rows) and measurements (bottom three rows) at 1\,mm PWV and boresight elevation of $60^\circ$ for the MF SATs. The top requirements are based on the map depths and observing efficiencies assumed in SO2019~\citep{SO2019_Forecasting}. The measured values, discussed in Section~\ref{sec:noise}, are the median telescope and effective per-detector NETs for observations around 1\,mm PWV with errorbars from the calibration uncertainty, including our best estimates of several sources of systematic uncertainty, at the end of commissioning.
}
\label{tab:requirements}
\setlength{\tabcolsep}{15pt}
\tablewidth{\linewidth}
\tablecolumns{7}
\tablehead{ 
    & \multicolumn{2}{c}{Two SATs} & \multicolumn{2}{c}{One SAT} & \multicolumn{2}{c}{Per Detector} \\
    & \multicolumn{2}{c}{$\si{\micro\kelvin}\mathrm{_{CMB}\sqrt{s}}$} & \multicolumn{2}{c}{$\si{\micro\kelvin}\mathrm{_{CMB}\sqrt{s}}$} & \multicolumn{2}{c}{$\si{\micro\kelvin}\mathrm{_{CMB}\sqrt{s}}$} \\
    \colhead{} & \colhead{f090} & \colhead{f150} & \colhead{f090} & \colhead{f150} & \colhead{f090} & \colhead{f150}
}
\startdata
\underline{Requirements} & & & & & & \\
Baseline NET & 3.4 & 4.3 & 4.8 & 6.1 & 312 & 395 \\
Goal NET     & 2.4 & 2.7 & 3.4 & 3.8 & 236 & 263 \\
\hline
\underline{Achieved} &  \\
SAT1 and SAT3 & $2.4\pm0.3$ & $3.2\pm0.4$ & & & & \\
SAT1 & & & $4.5\pm0.6$ & $5.4\pm0.8$ & $230\pm30$ & $320\pm50$ \\
SAT3 & & & $2.8\pm0.4$ & $3.9\pm0.5$ & $180\pm30$ & $260\pm30$ \\
\enddata
\end{deluxetable*}

The primary focus of the analysis in this paper is to demonstrate that the integrated telescopes meet or exceed the baseline instantaneous white noise levels assumed in SO2019, which presented baseline and goal map depths: 2.6 and 1.9\,$\si{\micro\kelvin}\mathrm{\,arcmin}$ for the f090 band and 3.3 and 2.1\,$\si{\micro\kelvin}\mathrm{\,arcmin}$ for the f150 band. These depths assumed a nominal five year survey covering 10\% of the sky with 20\% of the calendar time used in the maps. The SO2019 baseline and goal map depths were described as representing ``modest" and ``aggressive" technical developments, respectively, over experiments that were deployed at the time.

The top two rows of Table~\ref{tab:requirements} show the instantaneous sensitivity levels required to produce the SO2019 map depths with the SO2019 observing assumptions. Sensitivities are listed in terms of noise equivalent temperature in units of $\si{\micro\kelvin}\mathrm{\sqrt{s}}$ at 1\,mm precipitable water vapor (PWV).  The  ``Two SAT'' columns list the required combined inverse-variance weighted sensitivity of two MF SATs.  The baseline ``Two SAT'' instantaneous array noise values are the central requirements to be demonstrated during SO MF SAT commissioning. 

For analysis of the performance of the individual telescopes, we compare to the ``One SAT'' NETs, which are calculated assuming each telescope contributes equal weight to the ``Two SAT'' NET. Under the assumption that each detector contributes equally to the telescope sensitivity, the ``One SAT'' NETs can also be calculated as 
\begin{equation}
    \mathrm{NET_{telescope}^{CMB}} = \Gamma\left(\frac{\partial T_\mathrm{CMB}}{\partial T_{\mathrm{RJ}}}\right)
                                   \left(\frac{\partial T_\mathrm{RJ}}{\partial P}\right)
                                   \frac{\mathrm{NEP_{det}}}{\sqrt{2N_\mathrm{det}}}.
\label{eq:NET_telescope}
\end{equation}
\noindent In this equation, $\mathrm{NEP_{det}}$ is the per-detector noise equivalent power (NEP) at the detector in units of $\mathrm{aW/\sqrt{Hz}}$; the factor of $\sqrt{2}$ represents the conversion to $\mathrm{aW \sqrt{s}}$~(see eg, \cite{Zmuidzinas2003});  $N_\mathrm{det}$ is the number of operating detectors; $\partial T_\mathrm{RJ}/{\partial P}$ is a photometric calibration factor that converts changes in power on the detectors to an observed source temperature and $\partial T_\mathrm{CMB}/\partial T_{\mathrm{RJ}}$ converts that change into units corresponding to temperature fluctuations of the CMB blackbody.  Finally, $\Gamma$ is a factor that quantifies the level at which white noise levels are correlated between detectors \citep{bolocalc2018,Hill:24}.

We also calculate the associated required per-detector NETs 
as 
\begin{equation}
    \mathrm{NET_{det}^{CMB}} = \mathrm{NET_{telescope}^{CMB}}\Gamma^{-1}\sqrt{N_\mathrm{det}}.
\label{eq:per_det_NET}
\end{equation}

\noindent The values in the top two rows of Table~\ref{tab:requirements} assume 4243 (4816) live detectors corresponding to the baseline (goal) yield of 70\% (80\%) on a maximum of 6020 optical detectors. The per-detector numbers inherently include an averaged calibration of the telescope and are used to evaluate the optical and per-detector noise performance separately from the detector yields. Note that, throughout this paper, NET is the instantaneous sensitivity to an unpolarized input signal, which is approximately a factor of $\sqrt{2}$ smaller than the instantaneous sensitivity to input Stokes~$Q$ or $U$.\footnote{In general, this $\sqrt{2}$ factor should be multiplied by the polarization efficiency of each instrument. For a half-wave plate based experiment like the SO SATs, the polarization efficiency includes modulation efficiency of the half-wave plate.} 

The measurement of Equation~\ref{eq:NET_telescope} was chosen as the primary objective of commissioning because it establishes a quantitative picture of each telescope's optical performance, detector yield, and noise characteristics while simultaneously enabling comparisons to the white noise levels forecast by SO2019. Throughout this process, comparisons to contemporary instrument models were also performed to verify that the measured performance was consistent with design expectations. These models are implemented using the \texttt{jbolo} software \citep{Harrington_jbolo} and their details and implementation are discussed in Appendix~\ref{sec:jbolo}.

In the following sections: Section~\ref{sec:config} discusses the instrument configurations during commissioning, Section~\ref{sec:obs-summary} covers data taking for these measurements as well as the secondary and tertiary commissioning objectives, Section~\ref{sec:yields} discusses the focal plane performance with measurements of $\mathrm{NEP_{det}}$ and $N_\mathrm{det}$, Section~\ref{sec:cal} develops the calibrations for $\partial T_\mathrm{RJ}/{\partial P}$ and $\partial T_\mathrm{CMB}/\partial T_{\mathrm{RJ}}$ through planet observations, and Section~\ref{sec:noise} combines these results to demonstrate that both MF SATs meet or exceed the baseline instantaneous sensitivity requirements defined in SO2019.

\begin{figure}
    \centering
    \includegraphics[width=\linewidth]{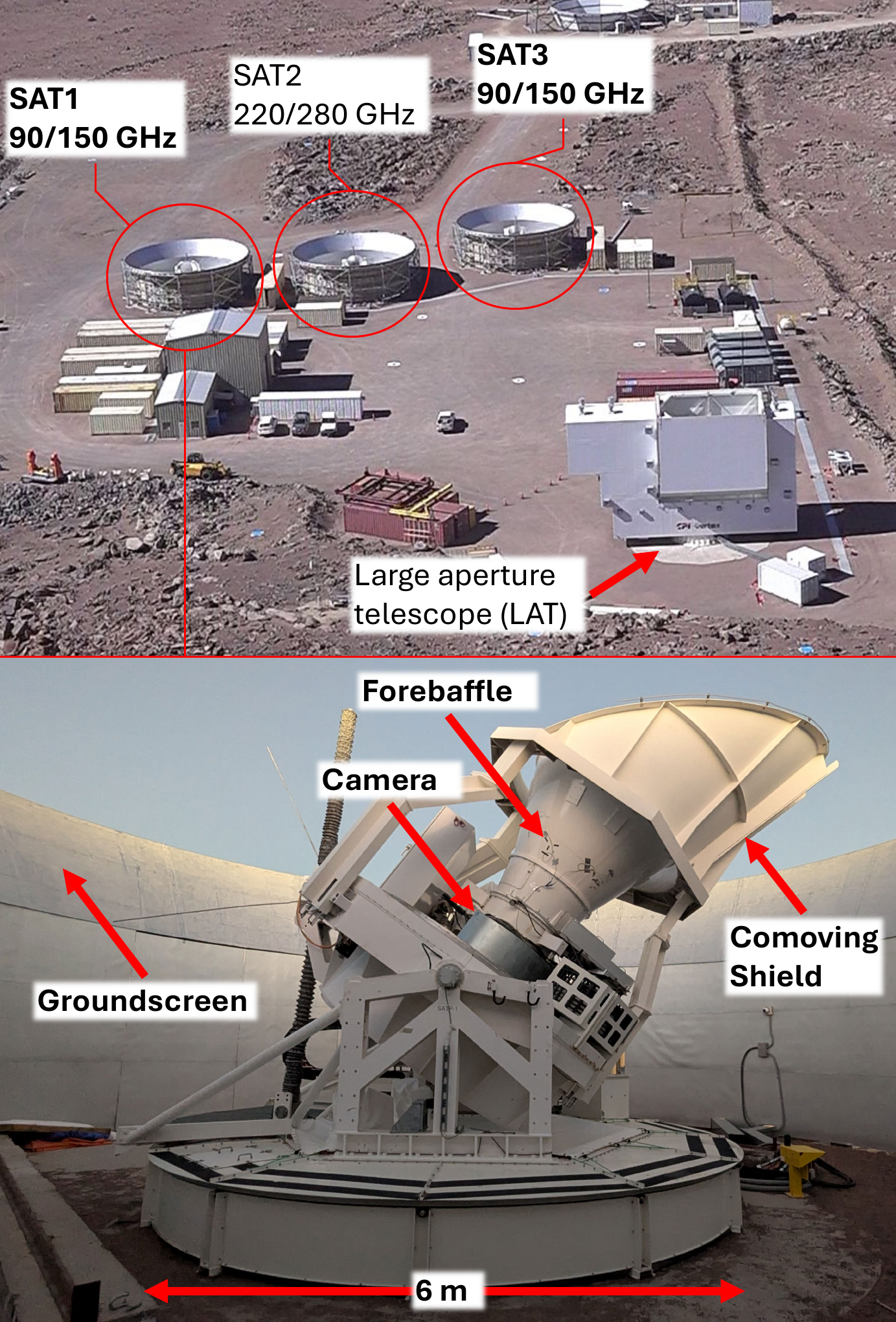}
    \caption{Top: Aerial image of the SO site as of December 2025 with the locations of the SATs and the LAT. Each SAT groundscreen is approximately 17\,m in diameter. Bold text indicates the location of instruments discussed in this paper. 
    Bottom: SAT1 viewed from inside the groundscreen with warm baffling components highlighted. The forebaffle has an absorbing (reflecting) internal surface for SAT1 (SAT3). The comoving shield is not installed on SAT3 for commissioning observations.
    }
    \label{fig:site}
\end{figure}

\section{Instrument Configurations \label{sec:config}}

Each SAT consists of a pointing platform that couples the cameras and optical baffling components, as shown in Figure \ref{fig:site} (see \citealt{Galitzki2024} for more details on the design and integration of the SATs). There are three SATs in the initial Simons Observatory configuration. SAT1 and SAT3 are the MF SATs are were both deployed to the SO site during the last quarter of 2023, with first-light observations occurring in October and December, respectively. 

The SAT1 hardware configuration maintains the ability to rotate the telescope in boresight, around the optical axis, and executes a scan strategy that includes daily rotations between $-45^\circ$, $0^\circ$, and $+45^\circ$. SAT3 was deployed without this ability as a cost- and time-saving measure. 
Systematics studies during the ISO time period will assess whether SAT3 should adopt a longer-term boresight rotation strategy, changing between boresight angles on a longer timescale and holding each one for an extended period rather than rotating daily.

A number of modifications to the two MF SATs have been made since the publication of \cite{Galitzki2024}. Here we detail the notable modifications to the instrument configurations that are reflected in the commissioning datasets.

\begin{deluxetable*}{lcccccc}
\tablecaption{A summary of the operating configurations and detector yields for the MF SATs during the commissioning time periods.
The on-site integrated yield is the number of readout channels regularly read out during operations and the number of detectors biasable at 1\,mm PWV are the number of detectors biased on transition for over half of the observations at elevation=$60^\circ$ and $0.8\leq \mathrm{PWV}\leq 1.2$\,mm. ``Good'' detectors are detectors that are biased on transition, and pass cuts on white noise and peak-to-peak timestream values as described in Section \ref{sec:yields}.
\label{tab:params}}
\tablecolumns{7}
\setlength{\tabcolsep}{15pt} 
\tablehead{
    \colhead{} &
    \multicolumn2c{SAT1 Pre-retrofit} &
    \multicolumn2c{SAT1} &
    \multicolumn2c{SAT3} \\
    \colhead{} &
    \colhead{f090} & \colhead{f150} &
    \colhead{f090} & \colhead{f150} &
    \colhead{f090} & \colhead{f150}
}
\startdata
Forebaffle                  & \multicolumn{2}{c}{Absorptive}
                            & \multicolumn{2}{c}{Absorptive}
                            & \multicolumn{2}{c}{Reflective} \\
Comoving Shield Installed   & \multicolumn{2}{c}{Yes}
                            & \multicolumn{2}{c}{Yes}
                            & \multicolumn{2}{c}{No} \\
Boresight Rotation          & \multicolumn{2}{c}{Yes}
                            & \multicolumn{2}{c}{Yes}
                            & \multicolumn{2}{c}{No} \\
Focal-plane Temperature Setpoint      & \multicolumn{2}{c}{85\,mK}
                            & \multicolumn{2}{c}{75\,mK}
                            & \multicolumn{2}{c}{62\,mK} \\
Commissioning Start         & \multicolumn{2}{c}{2024 May 25}
                            & \multicolumn{2}{c}{2025 Mar 10}
                            & \multicolumn{2}{c}{2024 Aug 1} \\
Commissioning End           & \multicolumn{2}{c}{2024 Dec 14}
                            & \multicolumn{2}{c}{2025 Jun 1 }
                            & \multicolumn{2}{c}{2024 Dec 14} \\
\hline
On-site Integrated Yield    & 4203 & 4012
                            & 5044 & 4848
                            & 4963 & 5072 \\
Biasable at 1\,mm PWV / sin($60^\circ$)       & 2578 & 3675
                            & 4728 & 4730
                            & 4711 & 4964 \\
``Good'' Detector Yield     & 2101 & 3491
                            & 4128 & 4348
                            & 4608 & 4743 \\
\enddata
\end{deluxetable*}

\subsection{Optical Components}

Several changes were made to the optical filter stacks within the telescopes. The most skyward filter for both SATs is now a polystyrene foam, radio-transparent multi-layer insulation (RT-MLI) filter \citep{Choi_2013}. As described in more detail in \cite{Day-Weiss_2026}, these filters were found to meet our optical and thermal requirements in the MF bands. 
We also replaced the original SAT3 40\,K alumina filter, which featured a meta-material anti-reflective surface, with one employing a two-layer mullite–duroid anti-reflective coating~\citep{Sakaguri2024}. This replacement is identical to that used in SAT1 and the change was made after identifying a defect in the anti-reflective surface that produced significant temperature-to-polarization leakage upstream of the half-wave plate. Data presented here are from observations after the filters were replaced.

\subsection{Warm Optical Shielding}
\label{subsec:warm_shielding}
There are three layers of warm optical shielding for the SATs; a conical ``forebaffle'' mated directly to the camera window, a reflective conic section ``comoving shield'' attached to the elevation structure, and a reflective ``ground shield'' built around the telescope platforms (see Figure~\ref{fig:site}). The combination of the comoving shield and ground shield was designed to satisfy the double-diffraction criteria for all terrain $<5^\circ$ above the horizon when the telescopes are observing at $50^\circ$ elevation (more details in \citealt{Galitzki2024}). The double-diffraction criteria ensures that any light from the ground must diffract at least twice off of optical shielding elements before entering the window aperture. 

During commissioning and ISO, two modifications to the design implementations were examined. First, the nominal observing elevation was raised to $60^\circ$ on both platforms which allows just the combination of the forebaffle and ground shield to meet the diffraction criteria, reducing the need of the comoving shield. The comoving shield was not installed on SAT3 to examine its effect on systematics. Second, modifications to the surface preparation of the conical forebaffle were studied.  The forebaffle extends out from the window and prevents geometric rays with $>40^\circ$ incidence angle relative to the optical axis from directly illuminating the window. 
SAT3 utilizes a bare aluminum reflective surface, while SAT1 operates with an absorptive forebaffle, where a layer of HR-10\footnote{\url{https://www.laird.com/products/absorbers/microwave-absorbing-foams/single-layer-foams/eccosorb-hr}} is attached to the aluminum forebaffle surface with a layer of Volara\footnote{\url{https://www.sekisuivoltek.com/volara}} on the exterior to provide weather sealing. 

During commissioning we established that the reflective forebaffle led to lower optical loading, and therefore lower photon noise on the detectors. However, based on experiences of other instruments, which have determined a preference for absorbing forebaffles to mitigate polarized signal leakage (e.g., \citealt{Li_2023}, \citealt{bicep2024_beam_calibration}), we chose to maintain the two different surface configurations throughout the ISO time period so a more comprehensive study of the systematic differences can be made. A summary of the different optical configurations is described in Table~\ref{tab:params}. Their longer term systematics evaluation will be carried out as part of the ISO analysis.

\subsection{SAT1 Focal Plane Retrofit}
\label{sec:retrofit}

Each SAT focal plane consists of seven universal focal-plane modules (UFMs) that house a detector wafer and associated 100\,mK microwave multiplexing readout components \citep{McCarrick_2021}. Several of the UFMs initially installed in SAT1 did not meet our performance requirements, mostly because they had overall lower integration yield than desired. While these modules were sufficient to measure observing conditions and telescope performance, four of the original seven were replaced for long-term observing. These modules were included in observations from first light through December of 2024, and are labeled ``SAT1 Pre-retrofit" throughout this paper. Two modules were replaced due to problems with the readout wiring, one had a broken flux ramp connection and a second had an open coplanar waveguide connection; both defects prevented half the module from being read out. The third module exhibited low optical efficiency in the f090 passband and the fourth had f090 detector saturation powers that were poorly matched to the realized optical loading during observations which resulted in high numbers of saturated detectors. 

Appendix~\ref{sec:preretro_data} discusses the NET performance of the pre-retrofit data. Of the yielded detectors, the per-detector performance in the pre-retrofit data is commensurate with the post-retrofit time period. The only exception to this is a lower optical efficiency f090 wafer. It is expected that data from this time period will be used in future cosmological analysis. 

During the retrofit, we also optimized the location of each module based on the saturation powers of the detector wafer and the measured optical loading from the different positions in the focal plane. Due to the wide $35^\circ$ field-of-view of the telescopes, there is a notable gradient in the atmospheric loading for detectors observing at higher elevations, and this motivated moving the wafers with lower saturation powers to these positions. Observations with SAT1 resumed in March of 2025, and the data from after this point are labeled ``SAT1'' in this analysis. Section~\ref{sec:yields} contains more details on the detector yields and which modules were replaced, installed, and moved. 

\subsection{Cryogenic performance}

The operating temperature for the MF SAT focal planes was designed to be 100\,mK. However, by operating the focal planes at lower temperatures, we can increase the saturation power of the detectors (see, for example, \citealt{Irwin2005}). This widens the range of PWVs where the detectors can operate. A heat strap in SAT1 was replaced with a higher thermal conductivity version during the focal plane retrofit. After this replacement, both SAT1 and SAT3 are capable of reaching base temperatures on the focal plane below 75\,mK. A setpoint control loop is used to servo the base temperature to a constant operational value. The operational temperature is chosen based on the individual instrument performance (see Table \ref{tab:params}). The systems have proven to be stable over long time periods with maintained cryogenic performance exceeding 280~days.

\begin{figure*}
    \centering
    \includegraphics[width=\linewidth]{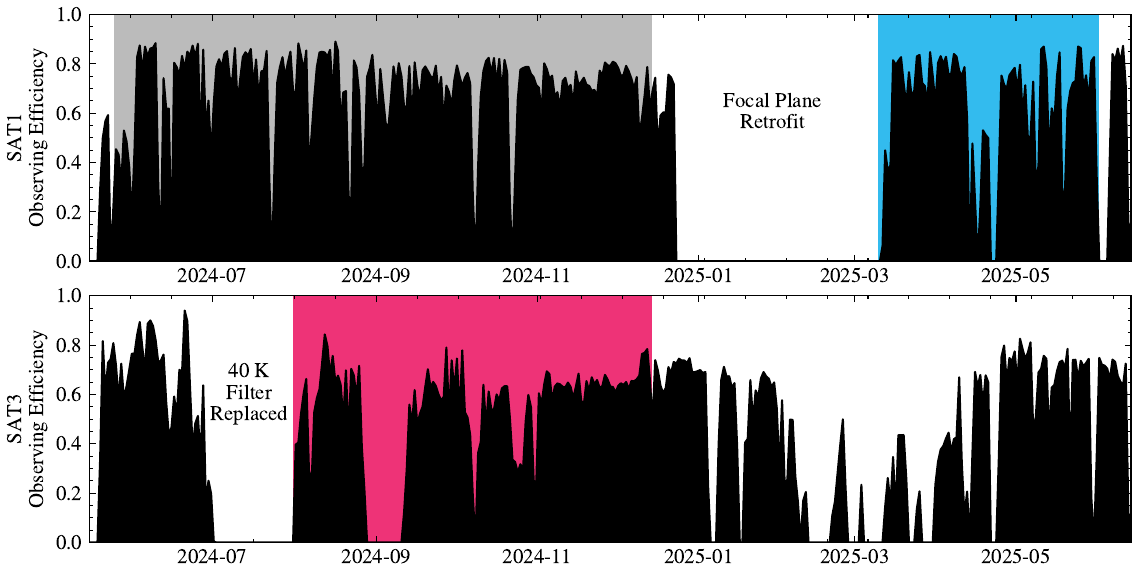}
    \caption{The realized observing efficiency, the black shaded regions show the fraction of each day SAT1 (top) and SAT3 (bottom) were taking data from either CMB or calibration observations from May 20th, 2024 until June 15th, 2025. The \textcolor[HTML]{808080}{\textbf{gray}}, \textcolor[HTML]{33BBEE}{\textbf{blue}}, and \textcolor[HTML]{EE3377}{\textbf{pink}} shaded regions denote the time periods used for the Pre-retrofit SAT1, SAT1, and SAT3 datasets, respectively.}
    \label{fig:obs_overview}
\end{figure*}

\section{Commissioning Observations}
\label{sec:obs-summary}

The work presented in this paper covers time periods from mid-2024 to mid-2025 from both SAT1 and SAT3 as listed in Table~\ref{tab:params} and shown in Figure~\ref{fig:obs_overview}. While commissioning and ISO had been planned as separate time periods, as realized, the observation time periods overlapped but had separate analysis objectives. For SAT3 we use the ISO time period that was set by the resumption of observations after the 40~K optical filter was replaced on August 1, 2024 and ended at the December 14, 2024 cutoff. The ``Pre-retrofit SAT1'' time period extends from when the comoving shield was installed on May 25, 2024 to the same December 14, 2024 cutoff. This dataset was used by the project as a test ISO period while anticipating the focal plane retrofit. The SAT1 dataset for this work spans about 3~months after the resumption of observations after the focal plane retrofit, from March 10, to June 1, 2025.

The initial stages of commissioning included a suite of measurements specifically targeted at developing the underlying instrument characterization necessary to perform the photometric calibrations discussed in Section~\ref{sec:cal}. First, Moon observations were used to determine the per-detector pointing offsets, to build telescope mount pointing models, and to establish a correspondence between the physical pixel locations of the detectors and the resonators used by the SLAC Microresonator RF (SMuRF) readout electronics~\citep{henderson/etal:2018, Yu_SMURF_2023, McCarrick_2021}. SMuRF uses microwave-frequency multiplexing to read out $\mathcal{O}(1000)$ detectors on a single readout channel, necessitating the development of techniques to correctly match tracked resonators to physical detectors~\citep{Lashner_2024}. The Moon is a bright $\sim210$\,K source with an angular diameter of $\sim30'$~\citep{Appel_2019}. The SO MF detectors saturate when directly viewing the Moon and its angular diameter is comparable to the on-sky pixel spacing of the SAT telescopes. This made it an ideal source to measure per-detector pointing information over a range of telescope elevation and boresight rotation angles.

Next, planet observations of Jupiter and Saturn were used to measure the per-wafer beam solid angles and the response to the planet. These sources are sufficiently bright, often available, and are effectively point sources for both the f090 and f150 frequency bands. The maximum apparent size of Jupiter during this time period is $0.042\,\si{\micro\steradian}$ which is $<1\%$ of the f150 beam solid angle. The signal-to-noise ratio of Jupiter is about ten times that of Saturn for the f090 and f150 bands, so Jupiter observations were prioritized over Saturn for the majority of the focal plane. However, Saturn is used to fill gaps in Jupiter coverage for upper portions of the focal plane where Jupiter was inaccessible due to a combination of Jupiter’s maximum daily elevation, platform limits on half-wave-plate operations at elevations below $48^\circ$,\footnote{This operating limit existed during the commissioning period but has since been changed through re-centering procedures implemented during subsequent cooldowns of the SAT cryostats. This has enabled more flexibility in targeting point sources.} and restricted boresight movement on SAT3. Planet observations are performed using constant-elevation scans that target one wafer in the telescope at a time. The per-wafer targeting optimizes the number of crossings for each detector so that reasonable beam maps are possible from single observations.

The beam products for SAT1 Pre-retrofit and SAT3 use planet observations from only the time ranges listed in Table~\ref{tab:params}. Additional observations from September through November 2025 are included for SAT1 to improve calibration precision to levels consistent with the other two periods. 
The Moon and planet observations were also used to rule out systematics, such as large near sidelobes or excess temperature-to-polarization leakage, that would have warranted hardware intervention.

Additional calibration measurements were taken during the commissioning time period to inform general operational procedures and scan strategy selection for the ISO period (commissioning goal 2). The cadence and method of performing detector and readout operations, such as re-biasing the TES detectors, were optimized to balance the tradeoff between downtime and sensitivity. A suite of azimuth scan speeds and turnaround accelerations were tested to search for and eliminate any that obviously impacted focal-plane thermal stability. Scan speeds and accelerations were chosen to minimize focal-plane heating during turnarounds while maximizing the efficiency of CMB and planet scans.  We also installed and performed the first calibration measurements with the sparse wire grid calibrators \citep{Nakata_2026} to verify the polarization response of the instrument. 

In between specific commissioning observations, the telescopes followed the prescribed CMB scan strategy to demonstrate operational readiness (commissioning goal 3). Figure~\ref{fig:obs_overview} shows the observing efficiencies of the two telescopes achieved from May 2024 until mid-June 2025 where the black shaded regions show the fraction of each day the telescopes were taking data from either CMB or calibration observations. The commissioning objective of $>50\%$ observing efficiency for more than three consecutive days was readily achieved. Longer operational downtimes were principally due to instrument updates or repairs while the shorter periods were usually due to inclement weather and Sun avoidance. The sustained observing efficiencies achieved by both SAT1 and SAT3 over this period demonstrate that the observatory is ready to operate the MF SATs at the levels required by SO2019 forecasts. Commissioning also validated the observatory control system software, site operations, and the infrastructure enabling remote observers to carry out telescope operations~(For more information on observatory control see \cite{Koopman_2024}, \cite{Bhimani_2024}, and \cite{Guan_2024}).

The CMB observations during the SAT1 Pre-retrofit, SAT1, and SAT3 included 3019, 984, and 1478 hours of data, respectively. For the yield and noise analysis during operations, approximately 1000 $\sim$1~hour long observations were randomly selected from the SAT1 Pre-retrofit and SAT3 datasets, while the entirety of the SAT1 observations were used. The top panel of Figure~\ref{fig:yields} shows the distribution of PWV measurements for the observations included in these datasets. As is typical for observations from Cerro Toco, the medians of these distributions are around 0.8\,mm of PWV and they have interquartile ranges from $\sim0.5-1.5$\,mm. The PWV measurement used in this analysis is always reported as the PWV at zenith as measured by a radiometer at the SO Site~\citep{pwv_monnitor_2011}. This radiometer has an operating range of $0.3-3$\,mm PWV so any observations below 0.3\,mm are set to that minimum value. Together, these observations form the datasets used in the following sections to characterize the detector yield and noise properties of the two MF SATs.


\section{Focal Plane Performance}
\label{sec:yields}

\begin{figure*}
    \centering
    \includegraphics[width=\textwidth]{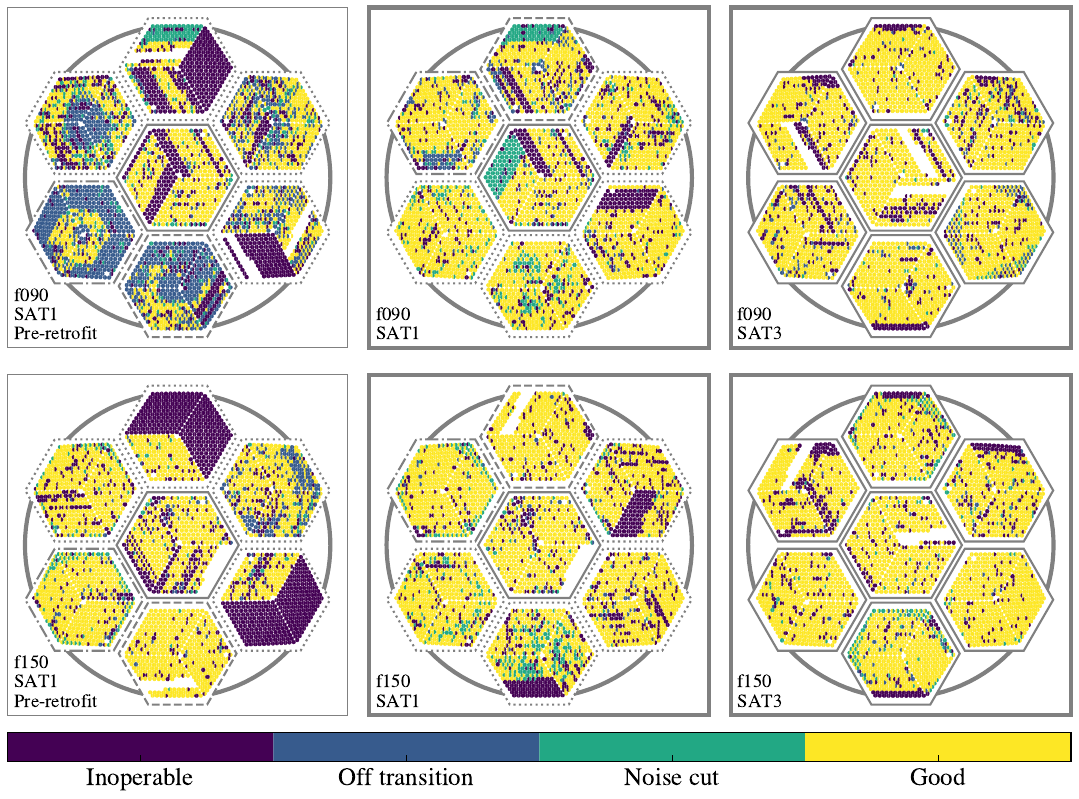}
    \caption{Schematic layouts of the SAT focal planes as projected onto the sky for a boresight angle of 0 deg, where the circles denote the 35 deg field of view of the instruments. Each column corresponds to a telescope and observing epoch and each row to a detector passband. For the SAT1 Pre-retrofit and SAT1 time periods, dotted wafer outlines denote UFMs that were removed and added, respectively. The dashed and dot-dashed wafer outlines are UFM-Mv18 and UFM-Mv22, respectively, these wafers were moved within the focal plane to reduce their on-sky optical loading. The colors in the plot mark the usual ``yield level'' for each detector. White areas within the hexagons indicate multiplexing chips that were not connected during wafer integration. ``Inoperable'' detectors were never read out or never matched to a resonator readout channel; entire inoperable rhombi were caused by broken flux ramp or coax connections to half a wafer while smaller sections arose from issues with multiplexing chips or bias line connections. 
    ``Off transition'' detectors were biased on transition at elevation=$60^\circ$ and PWV$ \in [0.75,1.25]$\,mm  less than half the time. For the SAT1 Pre-retrofit time period, these are primarily detectors that are saturated due to high optical loading. 
    ``Noise cut'' detectors were cut for noise values that were significantly higher than the rest of the wafer or were not well coupled to the atmosphere for more than half of the observations in the dataset. ``Good'' detectors pass all cuts more than half the time and their white noise measurements are used in the integrated telescope NET calculations in Section~\ref{sec:noise}.}
    \label{fig:wafers}
\end{figure*}

Each of the seven UFMs installed in a each SAT houses a detector wafer with 1720 optically coupled TES detectors \citep{Dutcher_2024}. Each MF UFM is identified by a unique label, ``MvX,'' where X is the same wafer number as used in \citet{Dutcher_2024}. Figure~\ref{fig:wafers} shows the layout of the focal planes for each telescope observing time period and the usual operating status, which will be discussed throughout this section, for each detector in the focal plane. Each column corresponds to one of the observing time periods and each row to a detector passband. The middle column in comparison to the left column shows the substantial yield improvement that was achieved with the SAT1 focal plane retrofit. Outlines on the wafers in this plot show which wafers were removed, added, or moved. Specific labels for which UFMs were installed in which positions are in Figure~\ref{fig:wafer_map}.

Each TES in a wafer is connected to one of twelve bias lines (six per passband) that are used to apply electrical bias power to the detectors as well as a microwave SQUID resonator that is coupled to one of two flux ramps and one of two coplanar waveguides. The flux-ramp and coplanar waveguide connections are used by the SMuRF system to read out the current going through the TES \citep{McCarrick_2021}. In addition, we use observations of the Moon to create a matching between detectors and resonator channels. Any broken bias line, flux ramp, or coplanar waveguide connection or lack of response to the Moon will result in detectors being labeled as ``inoperable'' as shown in Figure~\ref{fig:wafers}. Entire inoperable rhombi were caused by broken flux ramp or coax connections while smaller sections arose from issues with multiplexing chips or bias line connections. Table~\ref{tab:params} lists the on-site integrated yield for each telescope focal plane, which is the number of detectors that have passed all the initial integration cuts. SAT1 has 9,892 integrated detectors while SAT3 has 10,035, this represents 82\% and 83\% integration yield, respectively.

\begin{figure}
    \centering
    \includegraphics[width=\linewidth]{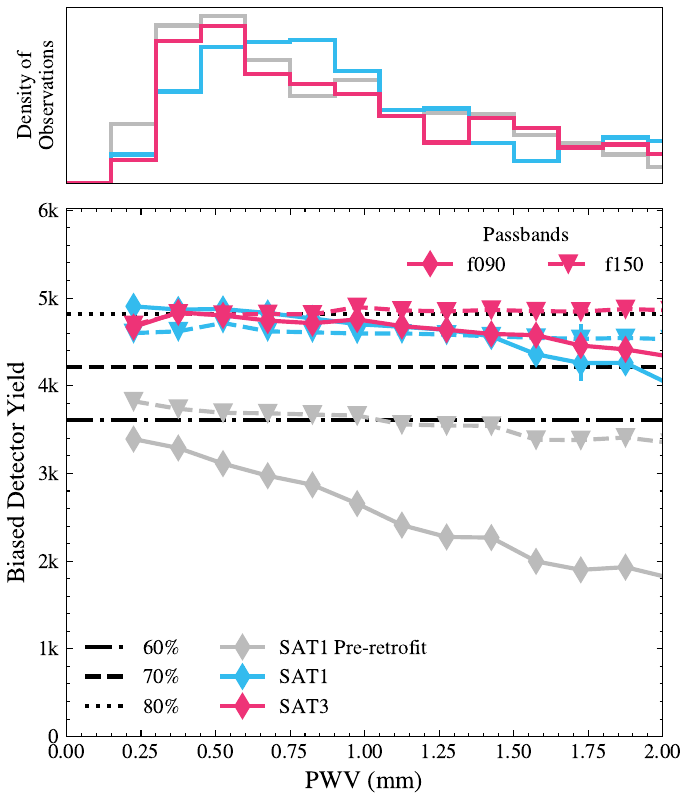}
    \caption{The median number of detectors biased on-transition with a fractional resistance between $0.2-0.8$ 
    for each PWV bin and each telescope observing epoch. 
    Fractional resistance is measured 
    via the bias step method. The histograms in the upper panel show the distribution of PWV values during each era as measured by the on-site PWV monitor. This monitor does not report values below 0.3\,mm. The SAT1 Pre-retrofit time period shows both the lower operating yields due to hardware and integration issues as well as detectors saturating as the PWV increases.
    Both SAT1 and SAT3 now operate with over 70\% of the detectors biased on transition at 1\,mm PWV and $60^\circ$~elevation, surpassing the forecasted baseline yield assumption.}
    \label{fig:yields}
\end{figure}

During operations, the detectors need to be biased ``on-transition'' in order to read out the optical power incident on each TES detector. This is accomplished using a two-step process. First, IV-load curves are taken where the voltage applied to each bias line is swept from high to low values to find the range of bias powers where the detectors transition from normal to superconducting. Second, for each bias line, a bias power is chosen to set as many detectors as possible on the middle of the transition curve. The power needed to bias the detectors decreases as the optical loading, the optical power incident on the detectors, increases \citep{Irwin2005}. The optical loading on the detectors arises from both instrumental and atmospheric emission and is therefore expected to vary with changing weather conditions. During commissioning and ISO, the biasing procedure is repeated every 4--5 hours during CMB scans and before every calibration observation. 

Bias steps are performed after the detectors are biased and between each CMB observation. Bias steps generate a two-point IV-curve, achieved by sending a square-wave bias voltage to the TES detectors \citep{Wang_2022}. From these, we derive the detector responsivity, electrical time constant, and TES resistance. Throughout this analysis, the detector responsivity is used to calibrate detector timestreams to $\SI{}{\pico\watt}$ and the electrical time constants, assumed to be equal to the optical time constants, are deconvolved from detector timestreams. The bias step operation calculates the ratio of the measured in-transition TES resistance to the normal resistance, $R_\mathrm{frac}=R_\mathrm{TES}/R_N$, where $R_N$ is measured by the preceding IV-load curve. We define detectors as on-transition if $R_\mathrm{frac} \in [0.2,0.8]$.  Figure~\ref{fig:yields} shows the number of detectors biased on-transition for each observing time period and each passband binned by PWV where each data point is the median value within the bin. In Figure~\ref{fig:wafers}, the ``off-transition'' label applied to detectors that were operable but found to be off-transition for more than half of observations when the PWV was between 0.75 and 1.25\,mm.

As discussed in Section~\ref{sec:retrofit}, a number of the SAT1 Pre-retrofit UFMs had low integration yield and had saturation powers in the f090 passband that were lower than desired when compared to the on-sky optical loading. This explains both the less than 70\% biased detector yield at low PWV and the significant drop off as the PWV increases. With the replacement detector wafers installed in SAT1, both telescopes now operate with 70--80\% of the detectors biased on transition at the operating elevation of $60^\circ$ for PWV values up to 2\,mm. 


Detector timestreams for each CMB observation are processed with a commissioning time-ordered data (TOD) pipeline to measure the white noise levels (NEPs) for power absorbed at each detector. The pipeline includes glitch and jump detection and repair, readout low-pass filter deconvolution, removal of the half-wave plate synchronous signal, and calibration to $\SI{}{\pico\watt}$ and deconvolution of the detector time constant using the most recent bias step responsivity measurements. The data are then demodulated at the half-wave plate modulation frequency ($\sim8$\,Hz) and an estimate of white noise is determined from $1/f$ plus white noise fits to power spectral density measurements of both the raw temperature timestreams and the demodulated $Q$ and $U$ timestreams. The noise fits are more stable for the demodulated timestreams and these timestreams more accurately represent the inputs to CMB map-making, so we report $\mathrm{NEP} = w_{n,Q}/\sqrt{2}$ where $w_{n,Q}$ is the white noise level fit from the $Q$ demodulated timestreams.\footnote{The power spectra of the demodulated timestreams result in more stable noise fits because, in comparison to the temperature timestreams, they are substantially less impacted by atmospheric $1/f$ variations and thus are much more white in the frequency range of interest. We verified that using the $U$ demodulated timestreams for this entire analysis produces the same results.}

The TOD analysis pipeline includes two additional sets of detector cuts: one applied to detectors with white noise levels exceeding $200~\mathrm{aW/\sqrt{Hz}}$ and another based on the distributions of peak-to-peak timestream values. The second cut is designed to remove detectors that are not well-coupled to the atmosphere. Detectors that pass these two additional cuts are labeled as ``good''  and used in the analysis of the focal plane performance shown here. In Figure~\ref{fig:wafers}, the ``good'' detectors are those that pass all cuts for over half of the analyzed observations for PWVs between 0.75 and 1.25\,mm. 

\begin{figure*}
    \centering
    \includegraphics[width=\textwidth]{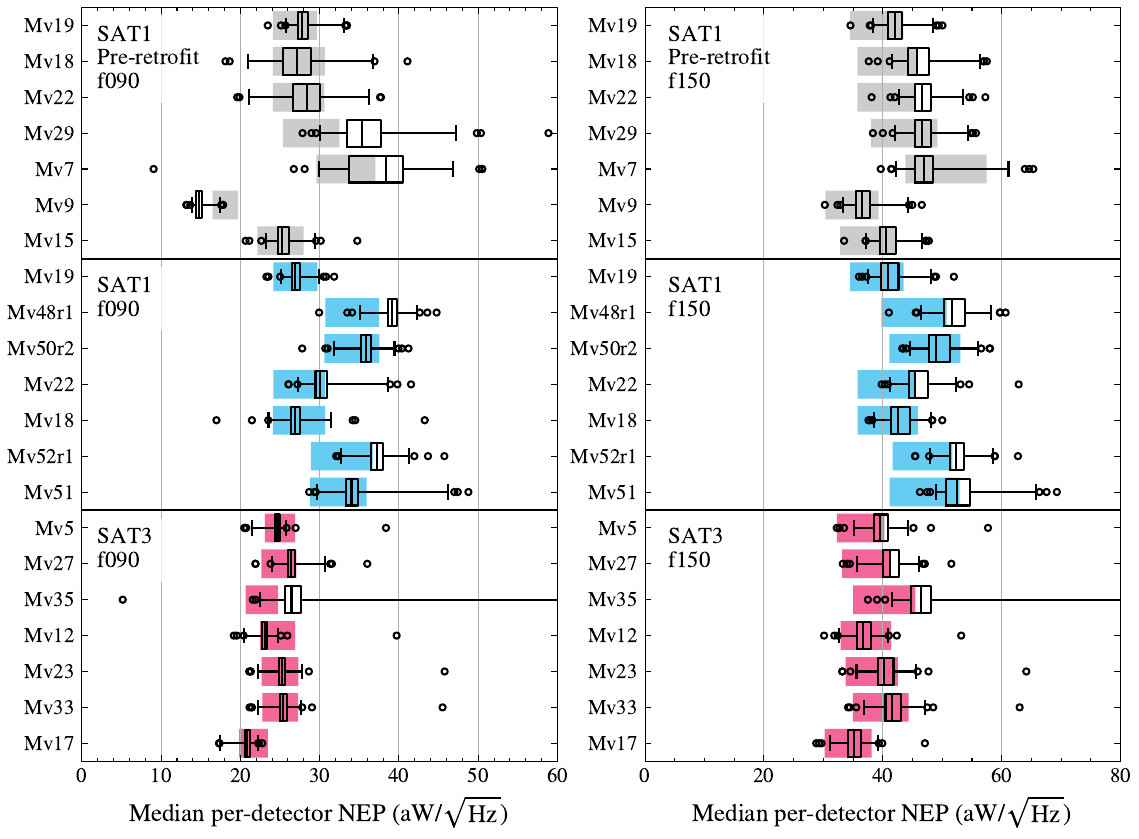}
    \caption{
    Box plots of the median per-detector NEP per-observation during the different observing eras, restricted to observations with PWV between 0.75 and 1.25\,mm at a boresight elevation of $60^\circ$. The NEP values are in \si{aW} on the detectors calculated using this bias step responsivity measurements discussed in the text. The width of the boxes encompass 90\% of the observations while the whiskers cover 1-99\% and dots indicate the remaining outlier observations. The colored shaded regions indicate per-wafer instrument model values for per-detector NEP that are built using detector and optics parameters measured during laboratory testing. 
    }
    \label{fig:det-neps}
\end{figure*}

The median NEP per detector, in $\mathrm{aW / \sqrt{Hz}}$, are shown in Figure~\ref{fig:det-neps} for observations with $0.75-1.25$\,mm PWV. The shaded regions of the plot are the outputs from a per-wafer instrument model that includes the median detector wafer parameters as measured by \cite{Dutcher_2024} and explained in more detail in Appendix~\ref{sec:jbolo}.\footnote{The SAT retrofit replacement wafers ( Mv48r2, Mv50r1, Mv51, and Mv52r1) were fabricated after the \cite{Dutcher_2024} proceeding and were made with higher targeted saturation powers than the original wafers.} These models account for each telescope's operating temperature and optical loading: SAT3, whose lower loading is partially attributed to its reflective forebaffle (see Section~\ref{subsec:warm_shielding}), has \loadingSATthree input at the feedhorns for the f090 (f150) bands, versus \loadingSATone for SAT1. The wafer-to-wafer variation in detector efficiency, saturation power, and critical temperature explains the majority of the intra-telescope differences in detector NEPs. The measured NEPs vary as a function of optical loading on the detectors and increase with PWV as would be expected as the atmospheric optical loading rises.

The measured NEPs for the majority of detector wafers fall within or near the ranges output by the instrument models. A few wafers show higher median NEP values than expected and these are caused by higher noise tails in the distribution that push the median values higher than the peak values. These noise tails were attributed to a higher sensitivity to mechanical vibrations in laboratory measurements, but have not been fully investigated. Integrated in the telescopes on-site, this effect is most noticeable in Mv7, Mv29, Mv48r1, and Mv52r1, where the median NEPs are 1--6~$\mathrm{aW/\sqrt{Hz}}$ higher than the peaks of the distributions. 
These per-detector NEPs, combined with the response calibration described in Section~\ref{sec:cal}, are used to derive the instantaneous telescope sensitivities presented Section~\ref{sec:noise}.




\section{Photometric Calibration and End-to-end Optical Efficiency}
\label{sec:cal}

A responsivity calibration of the instrument is required to convert detector white-noise levels into the NET statistics needed to assess the commissioning goals. The calibration developed here provides the conversion between the power incident on the detectors and the temperature of celestial point sources through planet observations. We also use this to measure the end-to-end efficiency for comparisons with expectations from the instrument models. While planet observations are sufficient for commissioning analyses, calibrations are expected to be determined through cross-correlation with \textit{Planck} observations for future cosmological analyses. 

Detector timestreams are calibrated into pW using bias step measurements and the per-detector responsivity is normalized per-wafer using a relative calibration factor generated from the per-detector coupling to atmospheric modes as described in \cite{Morris_2025}. Planet observations are integrated per-wafer into beam maps, which are then co-added across the focal plane to measure the beam solid angle ($\Omega_B$) and beam profiles for each telescope. We model the emission from planets as point sources with flux based on their angular diameter at the time of measurement and their temperatures based on \textit{Planck} measurements \citep{planck_LII_planet_flux} adjusted to the passbands of our telescopes. We use these to generate per-wafer photometric calibration factors. 

In principle, the beams and responsivity of the instrument can vary per detector. However, this analysis uses a single co-added beam across the whole focal plane and per-wafer response calibrations. The wafer-to-wafer variations in beam response are small enough to be wrapped into the uncertainty in the solid angle measurements here, but per-wafer beam products are being developed as more observations are taken. Per-wafer responsivity measurements are developed here because the wafer-to-wafer variations in detector parameters, including optical efficiency, are expected from in-lab testing.

\subsection{Beams and Solid Angle}
\label{sec:beams}

\begin{figure}[t]
    \centering
    \includegraphics[width=\linewidth]{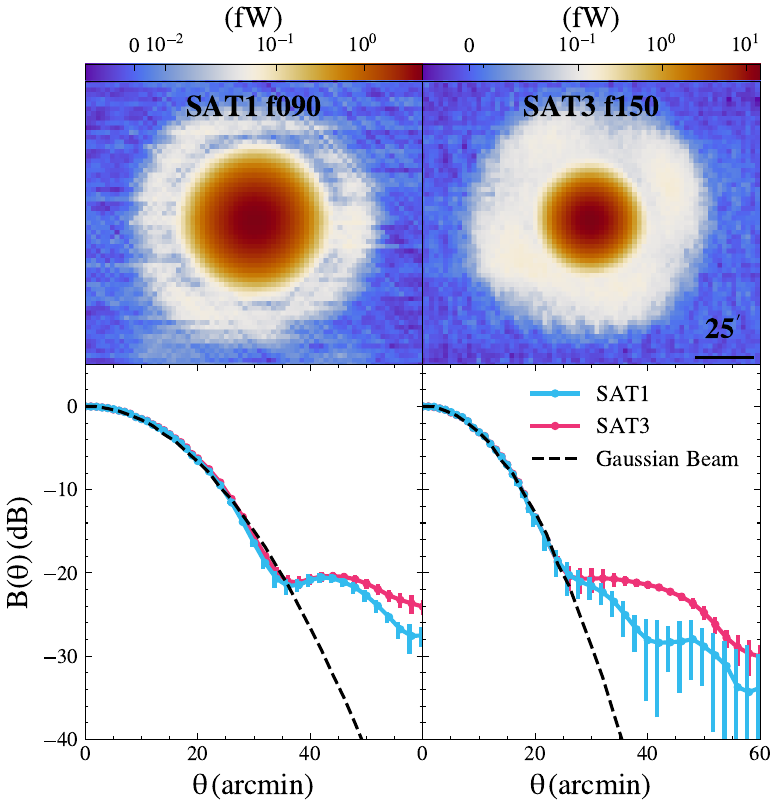}
    \caption{Co-added beam maps and beam profiles across the SAT1 and SAT3 focal planes. Top: The focal plane co-added beam maps for SAT1 f090 and SAT3 f150 in a $\mathrm{2^{\circ}\times2^{\circ}}$ area. Bottom: The azimuthally averaged beam profiles for the SAT1 and SAT3 observation epochs for both f090 and f150. The radial bin size increases from $1^\prime$ to $2^\prime$ at $\theta=20^\prime$ for easier viewing. Uncertainties are reported per radial bin as the standard deviation of the radial profiles for each detector module. A Gaussian beam is fit to the SAT1 (SAT3) focal plane co-added beam maps for the f090 (f150) passbands, and is shown as the dashed black line. The left (right) panel depicts the beam profile for the f090 (f150) detectors.}
    \label{fig:beams}
\end{figure}

The solid angle subtended by the telescope beam is integral to the development of the planet-based photometric calibration. In this section we use planet observations to measure radial beam profiles, $B(\theta)$, for each telescope and calculate their corresponding solid angles $\Omega_B$. 

From the planet observations discussed in Section~\ref{sec:obs-summary}, filter-and-bin maps were made from each observation using the principal component analysis (PCA) filtering technique described in \cite{Lungu_2022}. This method was optimized for the SO SATs using time-domain simulations of Jupiter by \cite{Dachlythra_2024}. We mask a region of $1^\circ$ radius around the planet, fill the masked region with a first-degree polynomial, calculate the dominant 40 modes from the PCA to be filtered out, unmask the planet signal, and finally remove the 40 PCA modes from the entire scan. 

Observations of an artificial source flown on a drone in the far field of the telescope~\citep{Coppi2025} identified a pointing deflection that rotates the detector pointing at the rotation frequency of the half-wave plate in both instruments. If unaccounted for, the pointing deflection would smear the beam and increase the full-width at half maximum (FWHM) and solid angle. We addressed this in SAT3 by implementing a time-dependent pointing deflection correction. The effect is small enough in SAT1 that it is not accounted for in this analysis. Details of this effect and further discussion on systematics of the main beams for both instruments are planned for future work.

The per-observation planet maps are all fit to 2D elliptical Gaussians and centered to correct for the fitted position offset. This centering is performed to remove residual telescope pointing errors. Before co-adding across observations, we apply four map selection criteria to the per-observation per-wafer beam maps. Maps are cut if their ellipticity, calculated as $\epsilon = (\sigma_x - \sigma_y)\,/\,(\sigma_x + \sigma_y)$ with $\sigma_{x,y}$ as the fitted beam widths of the major and minor axes, is greater than 0.04 and if their FWHM, calculated as $2\sigma_x\sqrt{2 \ln{2}}$, is more than $10\%$ greater than the per-observation median beam FWHM, $31'\,(22')$ for f090 (f150). These two cuts remove observations with poor data quality or source coverage. Third, we calculate the root mean square (RMS) error between the observed beam and the 2D Gaussian fit within $30'\,(20')$ for f090 (f150) of the beam center and cut beam maps with an $\rm{RMS} > 0.05\,\rm{pW}$. The RMS constraint is chosen to systematically cut maps with high noise.

Finally, a fourth cut utilizes the standard deviation in the peak height of 25 bootstrap map samples. We bootstrap each planet observation by creating 25 maps from the sample of $N_{\mathrm{det}}$ single-detector maps, where $N_{\mathrm{det}}$ is the number of detectors that contribute to the per-observation co-add. Each detector is chosen at random with replacement from the set of detectors that contribute to the co-add. These maps are also fit to 2D Gaussians and, if the standard deviation of the peak is greater than 15\% of the mean, the map is cut. This cuts observations with large variance in single detector maps. The combination of these selection criteria cuts poor quality maps while preserving enough observations for all detector wafers to have $\mathcal{O}(10)$ calibration measurements.

\begin{deluxetable}{lcc}
\tablecaption{A summary of the FWHMs, beam solid angles, and end-to-end optical efficiencies measured for the MF SATs during the commissioning time periods. The uncertainties labeled as statistical are the standard deviation of the values across the different wafers in the focal plane. Note, this is not strictly a statistical uncertainty as there expected differences in these parameters across the telescope focal plane. The systematic uncertainties are estimated based on input calibration uncertainies (ex. passbands for the optical efficiencies) and analysis choices (ex. beam profile fitting parameters). Note that the SAT1 Pre-retrofit optical efficiencies are reported without Mv9.
\label{tab:beamparams}}
\setlength{\tabcolsep}{30pt} 
\tablecolumns{3}
\tablehead{
    \colhead{} & \multicolumn2c{FWHM (arcminute)} \\
    \colhead{} & \colhead{f090} & \colhead{f150}
}
\startdata
SAT1 Pre-retrofit   & {\small $27.0\pm 0.4\;(stat)$}
                    & {\small $19.2\pm 0.2\;(stat)$} \\
SAT1                & {\small $26.9\pm 0.2\;(stat)$}
                    & {\small $19.3\pm 0.3\;(stat)$} \\
SAT3                & {\small $27.2\pm 0.2\;(stat)$}
                    & {\small $19.3\pm 0.2\;(stat)$} \\
\hline
                    & \multicolumn2c{Beam Solid Angle ($\mathrm{\mu sr}$)} \\
                    & f090 & f150 \\
\hline
SAT1 Pre-retrofit  
                    & {\small $80 \pm 6\;(stat) \pm 8\;(sys)$} 
                    & {\small $40 \pm 3\;(stat) \pm 4\;(sys)$} \\
SAT1                & {\small $77 \pm 3\;(stat) \pm 8\;(sys)$} 
                    & {\small $39 \pm 4\;(stat) \pm 4\;(sys)$} \\
SAT3                & {\small $84 \pm 4\;(stat) \pm 8\;(sys)$} 
                    & {\small $42 \pm 2\;(stat) \pm 4\;(sys)$} \\
\hline
                    & \multicolumn2c{End-to-End Optical Efficiency} \\
                    & f090 & f150 \\
\hline
SAT1 Pre-retrofit 
                & {\small $0.28 \pm 0.02\;(stat) \pm 0.05\;(sys)$} 
                & {\small $0.40 \pm 0.03\;(stat) \pm 0.08\;(sys)$} \\
SAT1            & {\small $0.31 \pm 0.02\;(stat) \pm 0.05\;(sys)$} 
                & {\small $0.40 \pm 0.03\;(stat) \pm 0.08\;(sys)$} \\
SAT3            & {\small $0.30 \pm 0.02\;(stat) \pm 0.05\;(sys)$} 
                & {\small $0.42 \pm 0.03\;(stat) \pm 0.08\;(sys)$} \\
\enddata
\end{deluxetable}

We next combine the per-observation maps into per-wafer planet maps by taking the weighted average of the dataset using the inverse square of the per-pixel noise as the weights. The per-wafer maps are also co-added into a per-telescope focal plane average, again using the inverse square of the per-pixel noise as weights. Azimuthally symmetric beam profiles,\footnote{The ellipticity is sufficiently small ($\epsilon<0.04$) such that we can take the azimuthal average of the beam without introducing a significant bias from the ellipticity of the beam.} $B(\theta)$, where $\theta$ is the angular distance from the beam center, are built for all the per-wafer maps and the per-telescope map for each telescope observing epoch using a two-part model. Inside $\theta=60'$, the beam maps are radially binned at $1'$ resolution and the binned data is interpolated between points using a cubic spline. Outside $\theta=60'$, the beam's asymptotic behavior is estimated with a $\mathrm{wing}=\alpha\,/\,\theta^3+\beta$ fit as motivated by \cite{Lungu_2022} and references therein. The wing is fit between $\theta=[55,90]'$ for both passbands. 

The per-telescope beam profiles per passband, as measured from the SAT1 and SAT3 time periods and the co-added beam maps for SAT1 f090 and SAT3 f150 are shown in Figure \ref{fig:beams}. The beam profiles for the SAT1 Pre-retrofit period are in agreement with the SAT1 period; as expected given that no optical components were modified during the retrofit. 
The radial profiles are binned at $1^\prime$ resolution up to $\theta=20^\prime$ and then the bin size is increased to $2^\prime$ for ease of viewing. The uncertainty per bin is calculated as the standard deviation across the per-wafer beam profiles. 

The comparison between SAT1 and SAT3 profiles shows that SAT3 has a slightly more complicated sidelobe structure: the f090 beams share a diffraction minimum at $\theta\sim35'$ but the SAT3 wing has a higher response than the SAT1 wing. Similarly, the SAT3 f150 profile plateaus at about $-20$\,dB for $\sim15'-20'$, in comparison to SAT1, before decaying. This difference between telescopes has also been seen in measurements with the drone source and is being investigated. Differences in anti-reflection coating methods on the half-wave plate may explain the variance.\footnote{We do not expect the forebaffle surface treatments to be the source of the beam profile differences since the forebaffles have $\sim40^\circ$ opening angles and these features are present much closer in at $\sim40'$.} However, as the beam profiles and corresponding window functions are well measured, the difference is not expected to impact cosmological science results.  

The beam profiles are used to calculate the total beam solid angle, $\Omega_{B}$, as
\begin{equation}\label{eq:solid_angle}
    \Omega_{B} = 2\pi \int _0^{\frac{{\pi}}{2}} B(\theta)\sin{\theta}d\theta.
\end{equation}
\noindent The combined $B(\theta)$ profiles for each telescope are integrated from the beam center to $90^\circ$ to fully account for the effect of the wing behavior on the solid angle. The integrated solid angles for each of the per-telescope beam profiles are reported in Table \ref{tab:beamparams}. Uncertainties that are labeled as statistical for the solid angles indicated the variation across the per-wafer beam profiles in the telescope, where the standard deviation across the profiles is used to calculate the variation of solid angles that are $\sim 4\,\si{\micro\steradian}$ and $\sim 3\,\si{\micro\steradian}$ for the f090 and f150 passbands, respectively. There are differences in beam solid angle expected across the focal plane from the optical design, so this type of uncertainty is not expected to integrate down with more measurements, but since a single solid angle measurement is used per telescope in Section~\ref{sec:abscal}, we include this variation in the uncertainty quantification.
 

The systematic uncertainties on the solid angles are estimated by calculating the solid angles across a grid of two different choices of analysis parameters: the resolution of the radial bins used to bin the beam maps and the range of angles used to fit the wing in the beam profile. The angular resolution of the bins is varied between $0.75'$ and $1.75'$ while the start of the wing fit is changed from $55'$ to $65'$. Across these analysis parameters the standard deviations of resulting solid angles are $\sim 1\,\si{\micro\steradian}$ and $\sim 0.4\,\si{\micro\steradian}$ for the f090 and f150 passbands, respectively. Finally, we add a 10\% systematic uncertainty, in quadrature, based on initial beam mapping systematic studies that have shown the solid angle is affected by the choice of the number of PCA modes removed and profile fitting function. This analysis will be expanded in future work.

Since the solid angles are calculated solely from planet observations focusing on the main beam and nearby sidelobes, they do not include any potential contributions to the sidelobes from wider angles. With their statistical and systematic uncertainties characterized above, the per-telescope solid angles are used directly in the photometric calibration in Section~\ref{sec:abscal}.

\subsection{Passband and Temperature Estimation}
\label{sec:passbands}

\begin{deluxetable}{lcc}
\tablecaption{Band-averaged quantities used to develop photometric calibrations of the instruments; their sources are discussed in detail in Section~\ref{sec:passbands}.
\label{tab:bandparams}}
\tablecolumns{3}
\tabletypesize{\small}
\tablehead{
    \colhead{} & \colhead{f090} & \colhead{f150}
}
\startdata
RJ Band Center              & $93\pm5$~GHz & $146\pm5$~GHz \\
Signal Bandwidth            & $30\pm4$~GHz & $32\pm5$~GHz \\
\shortstack[l]{RJ to CMB Conversion\\$(\partial T_{\mathrm{CMB}}/\partial T_{\mathrm{RJ}})$}
                            & $1.3\pm0.1$ & $1.7\pm0.1$ \\
\hline
Jupiter RJ Temperature & $170.0\pm0.8$~$\mathrm{K}$ & $170.0\pm0.9$~$\mathrm{K}$ \\
Saturn RJ Temperature  & $146\pm2$~$\mathrm{K}$ & $144\pm2$~$\mathrm{K}$ \\
\enddata
\end{deluxetable}

Commissioning observations and analysis were performed before any in-situ passband measurements were performed for the SATs, but an estimate of the passbands is necessary to complete a photometric calibration, measure the end-to-end optical efficiencies of the telescopes, and convert between Rayleigh-Jeans and CMB temperature units. 

For this analysis, we have chosen to use passbands measured on-site with a Fourier transform spectrometer (FTS) for the LAT telescope adjusted for known variations and changes based on the SAT optical design. All SO MF detector wafers to date, across the LAT and SATs, implement the same band-defining on-chip filter design~\citep{mccarrick2021_ufms}. In-lab LAT optics tube measurements have shown repeatable variations of shifts in the band edges across each detector wafer at the 3\% level~\citep{Sierra_2025} and the on-site measurements of seven LAT wafers has constrained the inter-wafer variation of the passbands to the 1--2\% level for the center frequencies and to the 3--5\% level for the bandwidths. It is expected that at least these levels of variations will also exist for the SAT detector wafers. The LAT measurements were corrected for the difference in frequency-dependent aperture stop efficiencies between the SAT and LAT optical designs. These aperture-stop corrected measurements are in general agreement with the in-lab SAT measurements~\citep{Seibert2023} but required substantially smaller neutral density filter corrections. The calculated values for RJ band-center and bandwidth are listed in Table~\ref{tab:bandparams}. The exact details of the passbands are the likely the largest unknown within this analysis and we have implemented conservatively larger uncertainties to attempt to compensate. 

For reference temperatures of Jupiter and Saturn, we use those presented in \citeauthor{planck_LII_planet_flux}~(\citeyear{planck_LII_planet_flux}; \textit{Planck} LII hereafter), adjusted to the SO detector passbands. \textit{Planck} LII reports thermodynamic temperatures at reference frequencies $\nu_c=100$ and 143~GHz that are relative to the CMB monopole. It also reports ESA models, $M_p(\nu)$, describing the spectral shape of the planet emission.  We normalize the model by the measured temperature as 
\begin{equation}
    T_P(\nu) = T_{P,\mathrm{meas}}(\nu_c) \frac{M_p(\nu)}{M_p(\nu_c)}
\end{equation}
\noindent and use $T_P(\nu)$ to calculate a frequency-dependent brightness temperature,
\begin{equation}
    T_B(\nu) = B_\nu\left(T_P(\nu)\right)\frac{c^2}{2 \nu^2 k_{\mathrm{B}}},
\end{equation}
\noindent where $B_\nu(T)$ is the spectral radiance of a blackbody, $k_{\mathrm{B}}$ is the Boltzmann constant and $c$ is the speed of light \citep{RybickiLightman}. We then define a band-averaged temperature, labeled ``RJ'' to make it distinct from the frequency dependent brightness temperature, as 

\begin{equation}
    T_{\mathrm{RJ}} \equiv \frac{1}{\Delta\nu} \int T_B(\nu) \tau(\nu) d\nu \quad \rm[K].
    \label{eq:planet_temp}
\end{equation}

\noindent Here, $\tau(\nu)$ is the peak-normalized passband of the SAT telescope, $\Delta\nu$ is the integrated bandwidth, and the calculated $T_{\mathrm{RJ}}$ values for Jupiter and Saturn are listed in Table~\ref{tab:bandparams} with uncertainties that include contributions from the \textit{Planck} calibration and from our passband estimation. The RJ temperature developed here can be converted to $K_\mathrm{CMB}$ following

\begin{equation}
    \frac{\partial T_{\mathrm{CMB}}}{\partial T_{\mathrm{RJ}}} = \frac{\int \tau(\nu) d\nu}{\int t'(\nu)\tau(\nu) d\nu}
\end{equation}

\noindent where

\begin{equation}
    t'(\nu) = \frac{x^2 e^x}{(e^x-1)^2} \quad \text{with} \quad x= \frac{h\nu}{k_{\mathrm{B}}T_{\mathrm{CMB}}}.
\end{equation}

Emission from Saturn requires additional consideration of the rings occulting the planet based on the opening angle of the ring with respect to the viewing angle. The Saturn ring opening angle was $|\theta|<5^\circ$ throughout the whole commissioning time period. To account for this, we use the Survey 2 measurement from \textit{Planck} LII that was from an opening angle of $\sim3^\circ$ and added additional uncertainty to the temperature to cover for range of temperatures expected during the measurement time periods. 

\begin{figure*}
    \centering
    \includegraphics[width=\textwidth]{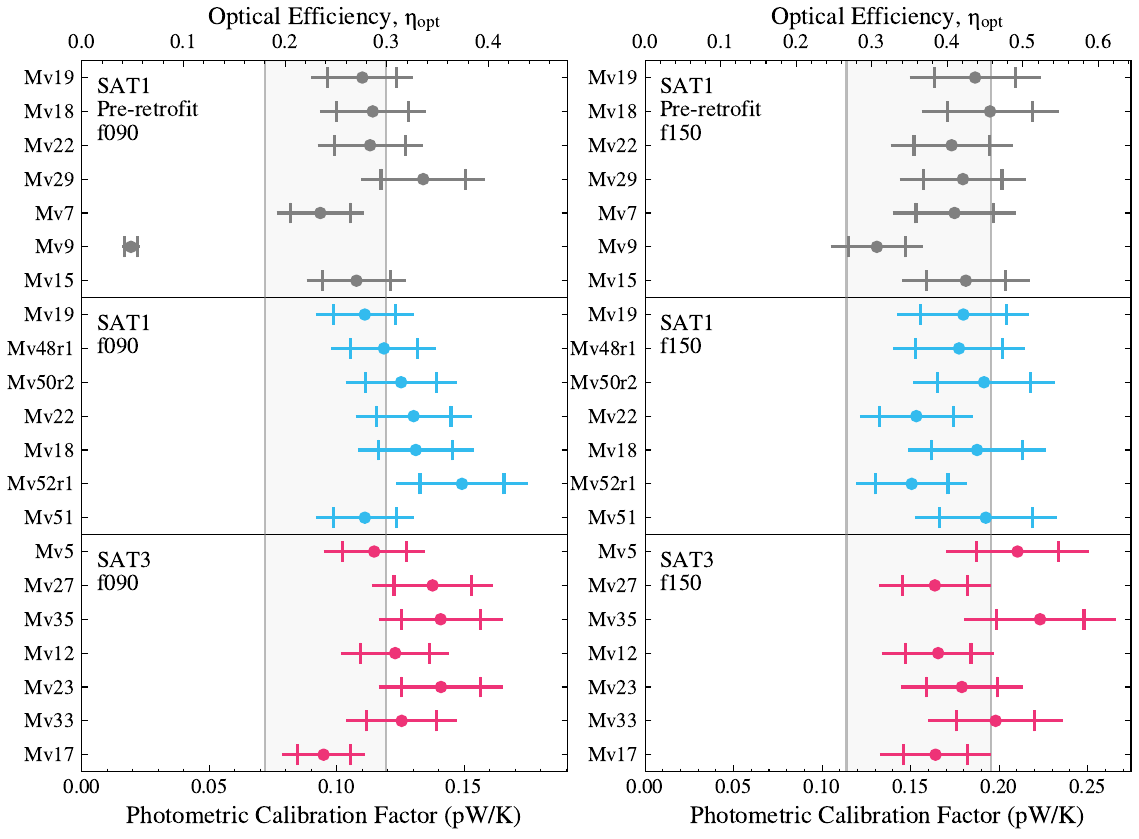}
    \caption{Per-wafer photometric calibration factors calculated from planet observations, normalized to include transmission through the atmosphere at 1\,mm PWV at an elevation of $60^\circ$. The uncertainties in per-wafer calibration factors are represented by the vertical lines off each data point. Derived optical efficiencies (Equation \ref{eq:e2e-eff}) are listed on the top $x$-axis. The uncertainties in the optical efficiency are larger than the photometric calibration, due to the extra dependence on bandwidth, and so these uncertainties are represented by the longer horizontal lines.  
    The f090 (f150) calibration factors are shown in the left (right) plot. The shaded region represents the range of expectations in response output from the instrument model. The overlap between the data points and the shaded regions indicate the end to end optical efficiency of the telescopes are in general agreement with expectations from the instrument design.}
    \label{fig:abscal}
\end{figure*}

\subsection{Photometric Calibration}
\label{sec:abscal}

The responses of the instruments are measured using the peak power observed in each wafer per planet observation and the per-telescope solid angle measurements ($\Omega_{B}$). For single-moded single-polarization detectors, the peak power observed from a point source, $S(t)$ will be 

\begin{equation}
    S(t) = \eta(t) \iint  I_\nu(t, \theta, \phi) B(\theta, \phi) \tau(\nu) A_\mathrm{eff}(\nu) d\Omega d\nu
\end{equation}
\noindent where $I_\nu(t, \theta, \phi)$ is the specific intensity of the source at the time of the observation, $B(\theta, \phi)$ is the telescope beam, $\tau(\nu)$ is the detector passband used in Equation~\ref{eq:planet_temp} and $\eta(t)$ is the total effective optical efficiency to the detector that includes variable loss from the atmosphere. For single-moded detectors, the effective area is 
\begin{equation}
A_\mathrm{eff}(\nu) =  \frac{c^2}{\nu^2 \Omega_{B}}   
\end{equation}
\noindent with $\Omega_B$ being the same as in Equation~\ref{eq:solid_angle}, which is equivalent to 
\begin{equation}
    \Omega_{B} = \int B(\theta, \phi) d\Omega.
\end{equation}
\noindent Integrating over solid angle and frequency, the expected peak power on a detector is 

\begin{equation}
    S(t) = \eta_\mathrm{atm}(t)\eta_\mathrm{opt}  k_{\mathrm{B}}\Delta\nu \frac{\Omega_P(t)}{\Omega_B}  T_{\mathrm{RJ}} , 
\end{equation}
\noindent where $\Omega_{P}(t)$ is the solid angle subtended by the planet for each observation. The planet solid angles are estimated from the JPL Horizons ephemerides software \citep{Horizons}, and include the time-dependent distance and viewing angle oblateness correction used in \textit{Planck} LII.
We have also split the total optical efficiency into variable contributions from the atmosphere, $\eta_{\mathrm{atm}}(t)$, and a constant contribution from the instrument, $\eta_{\mathrm{opt}}$. These factors are used in Equation~\ref{eq:NET_telescope} as $\partial T_\mathrm{RJ}/\partial P \equiv T_\mathrm{RJ}/S(t)$.

Finally, we define a photometric calibration factor,  

\begin{equation}
    \mathcal{C} = \eta_\mathrm{atm}(a_n)\eta_\mathrm{opt} k_{\mathrm{B}}\Delta\nu \quad \rm[pW/K],
    \label{eq:cal_factor_def}
\end{equation}

\noindent where $a_n = (\mathrm{PWV = 1\,mm, elevation = 60}^\circ$) is defined as nominal atmospheric conditions to have a consistent point of comparison between observations and to compare to the instrument model, which defaults to the telescopes observing the sky at 1\,mm PWV at $60^\circ$ in elevation. The $\eta_\mathrm{atm}$ values are calculated using the \texttt{am} atmospheric modeling software \citep{paine2019am} for the Cerro Toco plateau across a range of observing elevations and PWVs. For observations at elevation $=60^\circ$ and PWV $\leq 2$\;mm, the ratio of $\eta_\mathrm{atm}(a_n)/\eta_\mathrm{atm}(a)$ varies from 0.99 to 1.01 for the f090 band and from 0.98 to 1.02 for the f150 band.

For each planet observation, $i$, we measure the response per-wafer as $S_{w,i} \pm \sigma_{i}$ where $S_{w,i}$ is the peak height from the 2D Gaussian fit discussed in Section~\ref{sec:beams} and $\sigma_{w,i}$ is the quadrature sum of the standard deviation of the peak heights in the bootstrap maps, the map RMS measured in an outer annulus around the main beam, and a 6\% bias step calibration uncertainty around $S_{w,i}$. To simultaneously account for the per-observation measurement uncertainties and the uncertainties in the planet temperatures, an orthogonal distance regression fit is performed for $\beta_w$ as

\begin{equation}
     S_{w,i}\left(\frac{\eta_\mathrm{atm}(a_n)}{\eta_\mathrm{atm}(a_i)}\right) =  \beta_{w}\;\Omega_{p,i}\;T_{\mathrm{RJ},i}
\end{equation}

\noindent where $\beta_{w} = \mathcal{C}_w / \Omega_B$. Finally, $\mathcal{C}_w$ is calculated using the $\Omega_B$ values for each telescope and passband with uncertainties calculated as the quadrature sum of the fit parameter uncertainty and the total uncertainty on the solid angle. The resulting uncertainties in calibration factors are dominated by the uncertainty in solid angle.

The average calibration factors and uncertainty per wafer and per passband are shown in Figure~\ref{fig:abscal}, where the bottom $x$-axis denotes the calibration factor values. Uncertainties in the average calibration factor are represented by the vertical errorbar ticks. The shaded region in the plot marks the expectations from the instrument model. As discussed more in Appendix~\ref{sec:jbolo}, the instrument model used in this comparison is built to represent the range of expected performance metrics from individual components of the telescope (e.g., the spread of detector efficiencies measured by \citealt{Dutcher_2024}) and so represents the range of possible expected outcomes of the calibration measurement.

Finally, we estimate the end-to-end optical efficiencies per wafer. Following from Equation~\ref{eq:cal_factor_def}, the optical efficiency of the wafers is

\begin{equation}
    \eta_{w,\mathrm{opt}} =\frac{\mathcal{C}_w}{k_{\mathrm{B}} \Delta\nu \eta_\mathrm{atm}(a_n)}.
    \label{eq:e2e-eff}
\end{equation}

\noindent The end-to-end optical efficiencies derived from the calibration factors are shown in Figure~\ref{fig:abscal} on the top $x$-axis. The uncertainties on the optical efficiency include all the sources of uncertainty for the photometric calibration factors as well as uncertainty on the assumed bandwidth of the detectors. This larger uncertainty is indicated by the longer horizontal line on each data point in Figure~\ref{fig:abscal}. Based on the telescope optics design and our understanding of the optical elements in the telescopes, the instrument model predicts the optical efficiencies to be within \etafninety for the f090 passband and \etafonefifty for the f150 passband. The telescope averaged efficiencies, listed in Table~\ref{tab:beamparams}, are $\sim0.3$ and $\sim0.4$, in sufficient agreement with expectations to verify the optics and detectors are preforming within commissioning tolerances and that the instrument model is a reasonable representation of the built telescopes. However, there are also hints that the measured optical efficiency, especially for the f090 passband, is higher than would be expected once wafer-specific parameters from in-lab testing are incorporated into the model (See Appendix~\ref{sec:jbolo}). Future analysis and additional calibration measurements, such as in-situ passband characterization, will help separate if this a measurement systematic or an underestimation in the instrument model. 


\begin{figure*}
    \centering
    \includegraphics[width=\linewidth]{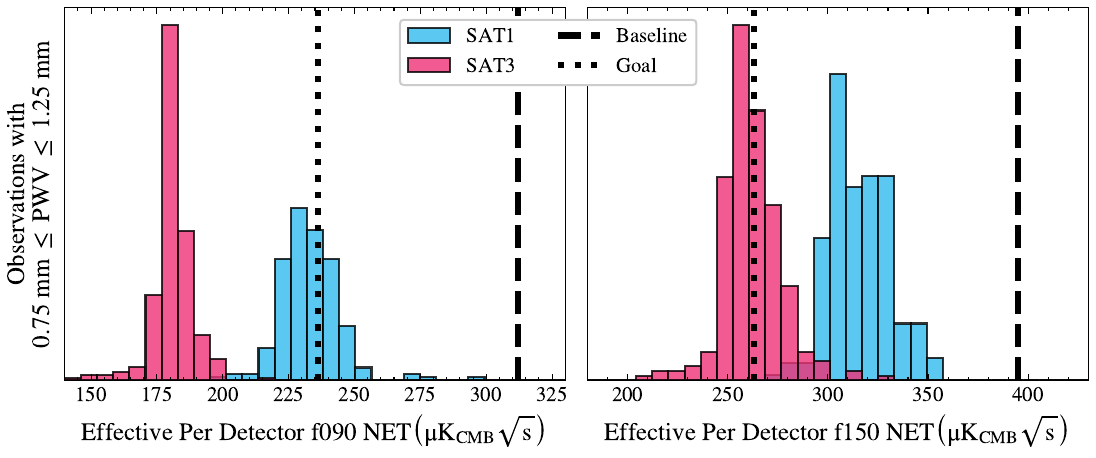}
    \caption{The instantaneous effective per-detector sensitivity of the two MF SATs after the SAT1 focal plane retrofit, for observations with PWV in the range of [0.75,1.25]\,mm. In all cases the per-detector performance of the instruments outperform the baseline sensitivity values, with SAT3 f090 well outperforming goal. As the end-to-end efficiencies of the instruments are roughly equal, the difference in performance between the two telescopes is primarily due to differences in the NEPs of the detectors. These are discussed in Section~\ref{sec:yields}.}
    \label{fig:net_per_det}
\end{figure*}

\section{Instantaneous Sensitivity}
\label{sec:noise}

In this section we determine the per-detector, per-telescope, and combined NETs for the MF SATs and compare them to the goals for commissioning. First, we define the per-wafer noise equivalent power, $\mathrm{NEP_{wafer}}$, using the per-detector NEPs and detector selection from the datasets in Section~\ref{sec:yields}, as

\begin{equation}
\mathrm{NEP_{wafer}} = \left(\sum_\mathrm{dets}  \frac{1}{(\epsilon_\mathrm{det}\mathrm{NEP_{det})^2}}\right)^{-1/2}.
\end{equation}

\noindent Here, the sum is over all detectors passing cuts and labeled as good in that pipeline, and $\epsilon_\mathrm{det}$ is the relative efficiency for detectors in the wafer calculated from the response to the atmospheric common mode. The $\mathrm{NET_{wafer}}$ is calculated from the $\mathrm{NEP_{wafer}}$ using the per-wafer calibration factors determined through planet observations in Section~\ref{sec:abscal}, $C_{w}$, and converted to CMB temperature units following the unit conversion from Section~\ref{sec:passbands}. The per-wafer $\mathrm{NET_{wafer}}$, corrected for atmospheric conditions at the time of measurement, is

\begin{equation}
    \mathrm{NET^{CMB}_{wafer}} = \Gamma\left(
    \frac{\partial T_{\mathrm{CMB}}}{\partial T_{\mathrm{RJ}}}\right)
    \frac{\eta_\mathrm{atm}(a_n)}{\eta_\mathrm{atm}(a_i)}
    \frac{1}{\mathcal{C}_{w}} 
    \frac{\mathrm{NEP_{wafer}}}{\sqrt{2}}.
\end{equation}

\noindent This equation is analogous to Equation~\ref{eq:NET_telescope} for a single wafer while accounting for variations in detector and optical performance across the wafer. We use photon correlation factors calculated from the instrument model as $\Gamma = 1.02 \pm 0.02$ for both passbands. The integrated instantaneous sensitivity of the entire telescope at nominal observing conditions is then 

\begin{equation}
    \mathrm{NET^{CMB}_{telescope}} =  \left[\sum_{\mathrm{wafers}} \left( \mathrm{NET^{CMB}_{wafer}} \right) ^{-2}\right]^{-1/2}.
    \label{eq:wafer_combine}
\end{equation}

The instantaneous detector sensitivities are shown in Figure~\ref{fig:net_per_det}, where the ``effective'' per-detector NET is as defined in Equation~\ref{eq:per_det_NET}. This definition of effective per-detector NET accounts for any long, higher-noise tails in the distribution of noise measurements across a wafer and averages over the individual wafer differences. At PWVs around 1\,mm, the detectors in both telescopes outperform the baseline requirement listed in Table~\ref{tab:requirements}, indicating that the combination of realized optical efficiencies and NEP values are well within the desired range. The differences in the detector NET performance between the telescopes is attributed to the spread of intrinsic detector wafer parameters and differences in optical loading on the detectors (See Sections~\ref{subsec:warm_shielding} and~\ref{sec:yields}).

\begin{figure*}
    \centering
    \includegraphics[width=\linewidth]{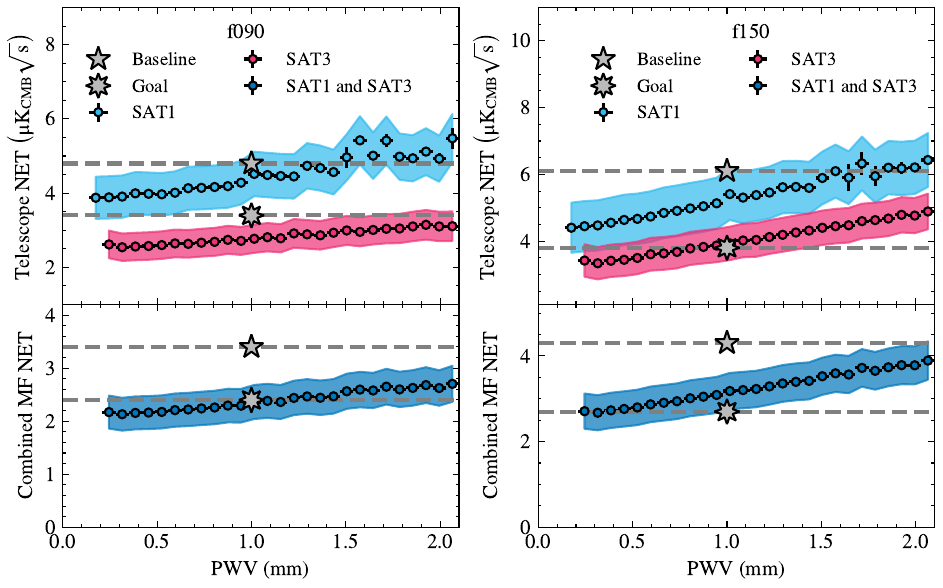}
    \caption{The instantaneous sensitivity of the two Simons Observatory MF SATs after the SAT1 focal plane retrofit individually (top) and combined (bottom). The f090 bands are on the left while the f150 bands are on the right. Each data point indicates the median telescope NET in a PWV bin, and the errorbars are the bootstrapped error on that median. The shaded regions denote the commissioning calibration uncertainty that is correlated across all data points. The baseline and goal values, denoted with the dotted lines and stars, are from Table~\ref{tab:requirements} for single SATs at PWV~$=1$\,mm. All telescopes are consistent with or outperform baseline noise levels while SAT3 f090 outperforms goal. The combined performance of the two telescopes outperforms baseline in both bands.}
    \label{fig:net_retro_sats}
\end{figure*}

The instantaneous noise performance of SAT1 and SAT3 individually are shown as a function of PWV in the top panel of Figure~\ref{fig:net_retro_sats}. The data for each PWV bin are the median noise measured within the bin, and the error bars are the bootstrapped errors on the median. The shaded regions are used to indicate the overall calibration uncertainty including the uncertainty in the calibration factors as developed in Section~\ref{sec:cal} and the uncertainty in the conversion to CMB temperature units. This calibration uncertainty is systematic to the measurement of telescope NET and does not integrate down with the addition of more observations to the dataset.

The baseline and goal single-telescope NET requirements are specified at $\mathrm{PWV}=1$\,mm, indicated by the stars in Figure~\ref{fig:net_retro_sats}. SAT1 is at baseline performance for both the f090 and f150 bands while SAT3 exceeds goal in f090 and is near goal at f150. The differences between the per-detector performance in Figure~\ref{fig:net_per_det} and the integrated telescope sensitivities in Figure~\ref{fig:net_retro_sats}, when compared relative to the baseline and goal lines, are due to the detector yields and the smaller PWV bins. For example, SAT1 f090 detectors are individually within about 1\% of goal level for a PWV bin between 0.75 and 1.25\,mm PWV, but the 75\% yield at 1\,mm PWV brings the telescope performance to a level 33\% higher than goal. Appendix~\ref{sec:preretro_data} discusses the changes in these performance metrics before and after the SAT1 focal plane retrofit.

The combined instantaneous sensitivity of SAT1 and SAT3 together is shown as a function of PWV in the bottom panel Figure~\ref{fig:net_retro_sats}. This is the final measurement for the MF SAT commissioning and reflects the overall performance of the telescopes. To be conservative, when combining the noise levels from SAT1 and SAT3 we assume the uncertainties are fully correlated. This is because the majority of the uncertainties are from calibrations that would be systematic across both instruments. The combined performance at f090 is $2.4 \pm 0.3\;\si{\micro\kelvin}\mathrm{_{CMB}\sqrt{s}}$ at 1\,mm PWV, consistent with the  goal level of $2.4\;\si{\micro\kelvin}\mathrm{_{CMB}\sqrt{s}}$. The instantaneous sensitivity of both instruments at f150 is $3.2 \pm 0.4\;\si{\micro\kelvin}\mathrm{_{CMB}\sqrt{s}}$ at 1\,mm PWV, which is between the baseline and goal values. These results reflect our best knowledge of the instrument performance at of the end of commissioning. We emphasize that the values reported here may evolve as we learn more about the instrument through, for example, a direct calibration to Planck, in-situ measurements of the passbands, and a deeper mapping of the sidelobes and wings of the main beam.

These results indicate the MF SATs now outperform the baseline instantaneous sensitivity levels used in SO2019. This was the measurement objective chosen for commissioning because it established a quantitative understanding of each telescope's integrated performance. Realizing the full map depths and scientific targets forecast in SO2019 will require achieving additional longer-term performance metrics; these include the level of detector analysis cuts used in mapmaking, the level of residual $1/\ell$ noise in the CMB maps, and the observing efficiencies throughout the survey. Nevertheless, achieving these commissioning targets demonstrates that the MF SATs are performing as designed and represents an important step toward the survey's full sensitivity goals.

\section{Conclusions}\label{sec:conclusions}

The commissioning of the Simons Observatory MF SATs took place between May~2024 and June~2025 and has successfully demonstrated that the noise performance of the instruments exceeds the baseline instantaneous sensitivity requirements used to forecast the scientific returns of the SO SAT program in SO2019. SAT1 was initially deployed with several detector wafer modules that had low integration yield, low optical efficiency or lower-than-desired saturation powers when compared to the optical loading measured on the instrument. This resulted in detector loss due to saturated detectors and a retrofit was performed in January~2025 to replace four wafers with newer versions fabricated with higher saturation powers. After the SAT1 retrofit, both SAT1 and SAT3 can bias over 70\% of the detectors on the focal plane onto transition for PWV values below 2\,mm, representing the vast majority of observing conditions from the Cerro Toco site. 

Planet observations, primarily of Jupiter, were used to measure the telescope beams and calibrate the overall power to temperature conversions for the instruments. The main beam shapes and optical efficiencies generally match the expectations from models of the instrument and future work will expand this analysis to per-wafer levels and more completely quantify the effects of systematics. 

The NEPs of the detectors were combined with the planet-based telescope calibrations to estimate the NET in CMB temperature units of the detectors and telescopes. These measurements represent the central requirement for commissioning. Overall, the NETs for SAT1 are $20-30\%$~higher than for SAT3 due to the spread of intrinsic detector parameters and differences in optical loading between the telescopes. The optical loading differences in the instruments come primarily from the choice to run SAT1 with an absorptive baffle and SAT3 with a reflective baffle. This choice enables additional instrumental characterizations and systematics mitigation that will be investigated in future work. 

The combined NETs of the two MF telescopes are near the goal level in the f090 band and between baseline and goal in the f150. These results indicate that the SO MF SATs have the instantaneous sensitivity needed to achieve the map depths forecast in SO2019, assuming the observatory maintains the data quality and observing efficiencies required to convert this sensitivity into final map depth. The commissioning of the first two MF SATs marks a critical milestone in the Simons Observatory program, validating the performance of the integrated systems under operational conditions and paving the way for full science operations.

\begin{acknowledgments}

This work was funded by grants from the Simons Foundation [MPS-Observatory-00457687, B.K.], Simons Foundation International [SFI-MPS-Observatory-00007127, B.K.], and in part by the U.S. National Science Foundation (Award Number: 2153201). SO operates in the Parque Astronómico Atacama in northern Chile under the auspices of the Agencia Nacional de Investigación y Desarrollo (ANID). We thank the Republic of Chile for hosting SO in the northern Atacama, and the local indigenous Licanantay communities, whom we follow in observing and learning from the night sky. This work would not be possible without the Simons Observatory site team based in Chile.

We acknowledge the support of REUNA, AmLight, and Parque Astron\'omico in enabling the real-time transfer of data via a fiber-optic connection.  This research used resources of the National Energy Research Scientific Computing Center (NERSC), a Department of Energy User Facility (HEP project mp107 2023-2026). The work presented in this article was also performed on computational resources managed and supported by Princeton Research Computing, a consortium of groups including the Princeton Institute for Computational Science and Engineering (PICSciE) and Research Computing at Princeton University.

Argonne National Laboratory’s work was supported by the U.S. Department of Energy, Office of High Energy Physics, under contract DE-AC02-06CH11357. We gratefully acknowledge the computing resources provided on Bebop, a high-performance computing cluster operated by the Laboratory Computing Resource Center at Argonne National Laboratory.

S.~Aiola acknowledges the hospitality of the Physics and Astronomy Department at USC. M.L.~Brown acknowledges funding from the Science and Technology Facilities Council (STFC) (grant numbers ST/X006336/1 and ST/X006344/1). Y.~Chione acknowledges the support from JSPS KAKENHI Grant Number JP24K00667. N.~Dachlythra and F.~Nati acknowledge funding from the European Union (ERC, POLOCALC, 101096035). Views and opinions expressed are, however, those of the authors only and do not necessarily reflect those of the EU or the ERC. Neither the EU nor the granting authority can be held responsible for them. 
R.~Dunner thanks ANID for grants BASAL CATA FB210003, QUIMAL-240004 and FONDECYT-1262583. J.~Errard acknowledges the SCIPOL project funded by the European Research Council (ERC) under the European Union’s Horizon 2020 research and innovation program (Grant agreement No. 101044073). 
J.~E.~Gudmundsson gratefully acknowledges funding by the European Union (ERC, CMBeam, 101040169). and the University of Iceland Research Fund.
M.~Hasegawa acknowledges JSPS KAKENHI Grant Numbers, JP25H00403. 
A.~D.~Hincks acknowledges support from the Sutton Family Chair in Science, Christianity and Cultures, from the Faculty of Arts and Science, University of Toronto, and from the Natural Sciences and Engineering Research Council of Canada (NSERC) [RGPIN-2023-05014, DGECR-2023-00180].
R.~Hlo\v{z}ek acknowledges NSERC grants RGPIN-2025-06483 and SMFSU-60768. L.~Page acknowledges Wilkinson and Misrahi Research Funds. Y.~Sakurai acknowledges JSPS KAKENHI grant number: JP23H01202. O.~Tajima acknowledges JSPS KAKENHI grant numbers JP22H04913 and JP17H06134.  A.~Thomas acknowledges funding from U.S. National Science Foundation (Award Number:  2153201).

This document was prepared by Simons Observatory using the resources of the Fermi National Accelerator Laboratory (Fermilab), a U.S. Department of Energy, Office of Science, Office of High Energy Physics HEP User Facility. Fermilab is managed by Fermi Forward Discovery Group, LLC, acting under Contract No. 89243024CSC000002.

\texttt{Claude Code} was used to reorganize analysis scripts, write software documentation, and improve plot formatting for the results in this paper. \texttt{Claude} was also used for proof-reading the manuscript text.

\end{acknowledgments}


\software{
\texttt{astropy} \citep{2013A&A...558A..33A,2018AJ....156..123A,2022ApJ...935..167A}, 
\texttt{pyephem} \citep{PyEphem},
\texttt{am} \citep{paine2019am},
\texttt{jbolo} \citep{Harrington_jbolo},
Some of the software developed to produce this analysis can be found publicly available on the Simons Observatory GitHub organization (\url{https://github.com/simonsobs}). 
}


\appendix
\section{Instrument Modeling}
\label{sec:jbolo}


In addition to comparing to forecasting values, commissioning calibrations and noise measurements are also compared to an instrument model that is built off the current best understanding of the individual elements installed in the telescopes. These models are built using the \texttt{jbolo} software which takes input definitions about the telescope optics and detector parameters and calculates a variety of integrated performance metrics for a telescope, including instantaneous sensitivity~\citep{Harrington_jbolo}. SO has used models implemented in \texttt{jbolo}, or an earlier software iteration \texttt{BoloCalc}~\citep{bolocalc2018}, to track expected telescope performance since the initial instrument models for SO2019 were developed. The instrument models for this paper are available at \url{https://github.com/simonsobs/sat-mf-cmg-instrument-model/tree/main}.

As with any model, these are a simplified representation of the true system, and decisions have to be made for how to integrate the combined effects of variation of parameters across the focal plane and the remaining uncertainty in the various model inputs. We address this by producing two models, ``Upper'' and ``Lower,'' which are intended to bracket the expected performance.
For example, the loss tangent of the vacuum window material, high-density polyethylene (HDPE), can vary by temperature, batch, and preparation and we do not have loss measurements for our specific windows. To reflect this uncertainty, the Upper model implements a loss tangent that would lead to better performance ($1\times10^{-4}$) while the Lower model uses a more pessimistic value ($3.1\times10^{-4}$; e.g.~\citealt{Elwood2024, Lamb1996}). 

The instrument models also include as many measurements as are available for the individual telescope components. This includes transmission or reflection measurements of the various optical elements; many of which are in \cite{Galitzki2024}, \cite{Thomas2026_foam}, and \cite{Sugiyama_2024_HWP}. In addition, the pre-deployment detector characterization campaign reported the realized distributions of many detector parameters such as critical temperature, normal resistance, and saturation power at 100\,mK~\citep{Dutcher_2024}. These are included as ranges where the Upper (Lower) model uses the value that would result in better (worse) noise performance.

Despite this effort to bound the model with measured component properties, the Upper and Lower models are not guaranteed to bracket the true performance of the telescopes; not every parameter is independently measured, some elements are only characterized as samples or witness pieces rather than in situ. Additionally, in-lab measurements will inherently have their own uncertainties and systematics that may not be perfectly accounted for. As a result, when a measurement falls outside the modeled range, or agrees with the model less well than expected, it is generally not possible to say with confidence whether the discrepancy reflects a limitation of the model, an unmodeled effect, or an issue with the measurement itself. We treat the instrument model comparisons throughout this paper as a useful consistency check and diagnostic tool rather than as ground truth.

\begin{figure*}
    \centering
    \includegraphics[width=\linewidth]{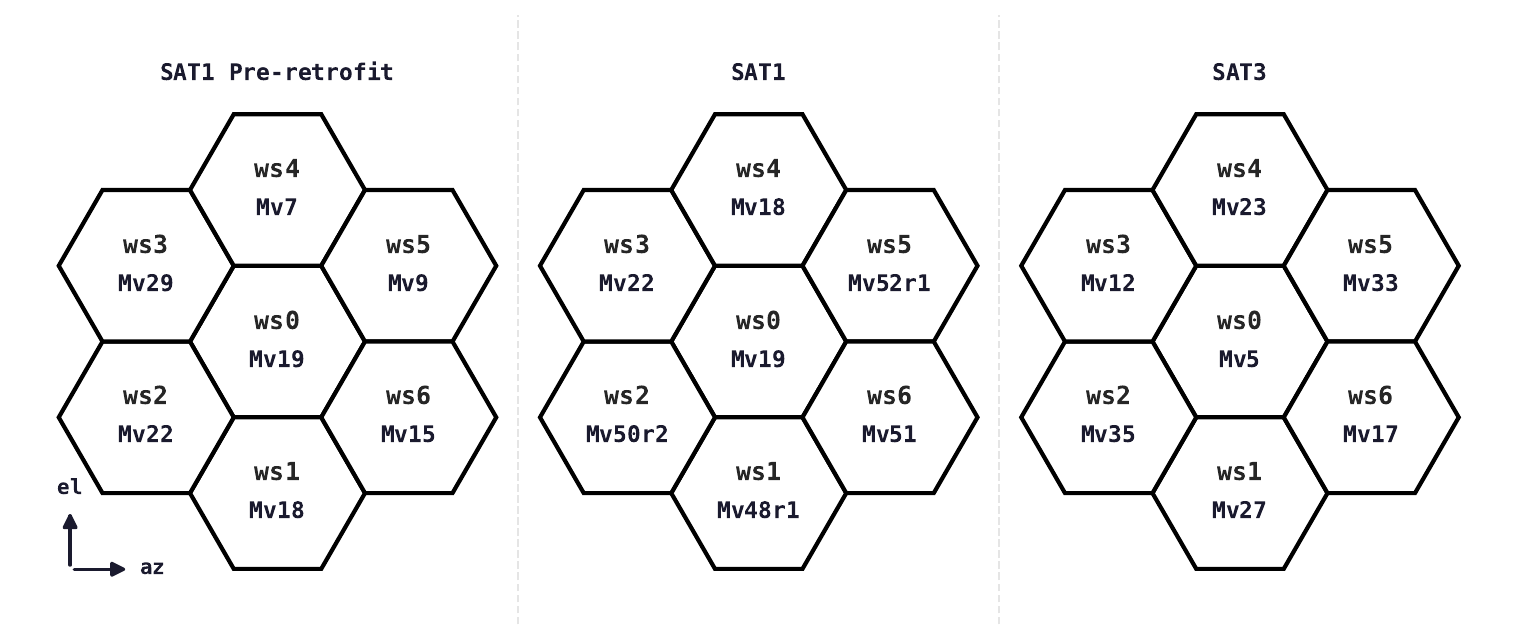}
    \caption{A map of where each UFM is installed in each SAT focal plane. Each hexagon represents a detector wafer as it would be projected on-sky with boresight roll = 0. The top ``wsX'' labels are the indexing of the wafer slot positions and the ``MvXY'' labels are the indexes of the UFMs that were installed in those slots for each telescope and time period.}
    \label{fig:wafer_map}
\end{figure*}

\begin{figure*}
    \includegraphics[width=\textwidth]{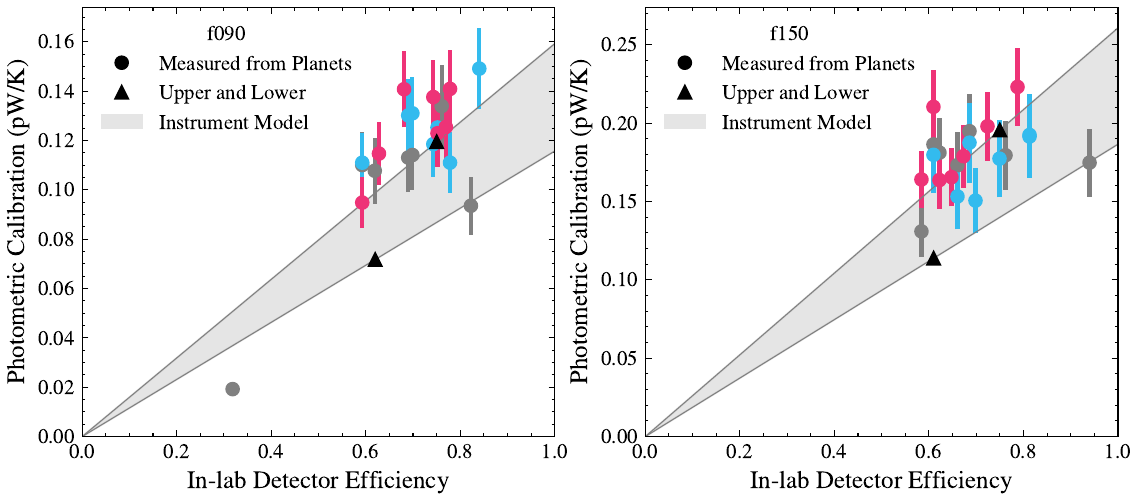}
    \caption{\label{fig:abscal-deteff} The measured photometric calibration factors for each wafer compared to their respective in-lab detector efficiency measurements for the f090 (left) and f150 (right) passbands. The marker colors denote which telescope each wafer is from following the standard color scheme in this paper where \textcolor[HTML]{808080}{\textbf{gray}}, \textcolor[HTML]{33BBEE}{\textbf{blue}}, and \textcolor[HTML]{EE3377}{\textbf{pink}} denote the Pre-retrofit SAT1, SAT1, and SAT3 datasets, respectively. The instrument model region is the range of expected performance between the Upper and Lower instrument models. The instrument model systematically under-predicts the calibration factors, especially for the f090 passbands, and investigation into the possible sources of the descrepancy will require calibration measurements beyond the scope of commissioning.
    The two Pre-retrofit SAT1 points outside the range of the instrument model are Mv9 and Mv7, that were both replaced during the focal plane retrofit.
    }
\end{figure*}

Interpreting the NEP values from each detector wafer, as shown in Figure~\ref{fig:det-neps}, required the development of more specific models that used the measured parameters for each individual wafer. To produce wafer specific models: each Upper and Lower model was augmented with wafer specific values for:
\begin{itemize}
    \item The focal plane temperature, $T_\mathrm{bath}$, for each telescope as in Table~\ref{tab:params}.
    \item Telescope specific spillover temperatures, 283\,K for SAT1 and 100\,K for SAT3, to account for the optical loading differences between the absorptive and reflective forebaffles.
    \item The in-lab measured detector efficiencies adjusted for updated knowledge of the detector passbands.
    \item The in-lab measured normal resistance, critical temperature $T_{\mathrm{c}}$, thermal conductance $G$ as well as $\kappa$ and $n$ such that $P_\mathrm{sat}(T_\mathrm{bath}) = \kappa(T_{\mathrm{c}}^n-T_{\mathrm{bath}}^n)$.
\end{itemize}

\noindent The map of which wafers were installed during each telescope obsering epoch is shown in Figure~\ref{fig:wafer_map}. The shaded regions in Figure~\ref{fig:det-neps} are then the range of predicted detector NEP from each wafer-specific model that includes contributions from photon noise, phonon noise, Johnson noise, and readout noise. These models are capable of recreating the measured median per-detector NEP for most wafers, with the exception of a subset of wafers that exhibit higher vibrational noise as discussed in Section~\ref{sec:yields}.

The instrument model expectation range in Figure~\ref{fig:abscal} is from the photometric calibration factor predicted by the general Upper and Lower models; and thus includes the range of performance possible from both the telescope optical elements and the distribution of measured detector efficiencies. Figure~\ref{fig:abscal-deteff} shows how these measurements align with the model as a function of detector efficiency and more clearly shows the measured photometic calibrations values are higher than expected, especially for the f090 band. This difference could be caused by one or a combination of possible sources: a systematic offset in the in-lab detector efficiency measurements, an underestimated aperture stop efficiency in the optics model, a wider than predicted passband, or some other unknown. Additional on-site measurements, particularly passband calibrations with an FTS and coherent beam maps using the calibration drone, will help distinquish between the possible sources of descrepancy. The two wafer-modules with substantially lower calibration factors, Mv7 and Mv9, were replaced as part of the SAT1 focal plane retrofit.

\begin{figure*}
    \includegraphics[width=\textwidth]{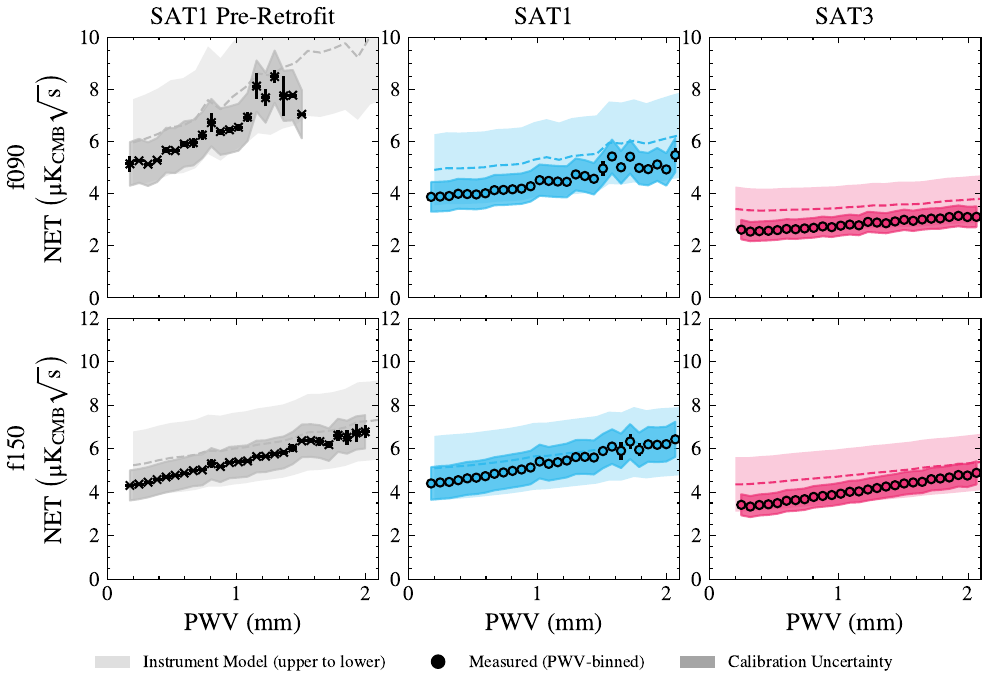}
    \caption{\label{fig:net_versus_model} The measured telecope NETs in comparisons to predictions from the instrument model after the achieved yield as a function of PWV is added into the instrument model.
    }
\end{figure*}

Figure~\ref{fig:net_versus_model} shows how the measured telescope NETs compare to the predictions from per-wafer instrument models. In this calculation, each of the per-wafer models is evaluated to predict the wafer integrated NET as a function of PWV with the yield in each wafer set to the median number of good detectors in that PWV bin. The per-wafer evaluations are then combined as in Equation~\ref{eq:wafer_combine} to produce the predicted telescope NET. As implemented, the instrument models slightly over-estimate the expected telescope NETs and this is likely related to the under-estimate of the optical efficiecy discussed earlier. The level of agreement between the model prediction and the realized performance highlights the substantial success for both the instrument model development and the pre-deployment in-lab characterization campaigns

\section{Effect of the SAT1 Retrofit}
\label{sec:preretro_data}

\begin{figure*}
    \centering
    \includegraphics[width=\linewidth]{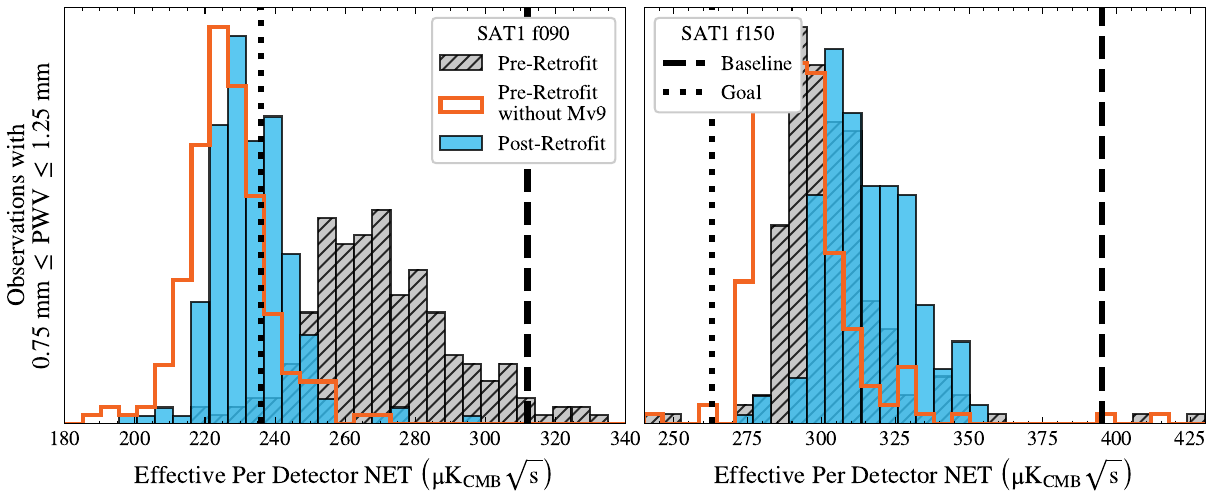}
    \caption{The instantaneous effective per-detector sensitivity for SAT1 before and after the focal plane retrofit for observations with PWV in the range of [0.75,1.25]\,mm. The pre-retrofit data is shown with and without the inclusion of UFM-Mv9, the detector wafer with lower optical efficiency in the f090 band. Excluding UFM-Mv9, the pre-retrofit SAT1 f090 per-detector NET is within 2\% of the goal value, validating the optical performance of the instrument; the modest increase in f150 per-detector NET after the retrofit is associated with the higher saturation powers of the replacement wafers.
    }
    \label{fig:net_per_det_all}
\end{figure*}

This appendix discusses the SAT1 instantaneous sensitivity before and after the focal plane retrofit. The Pre-retrofit per-detector NETs in Figure~\ref{fig:net_per_det_all} demonstrate that the individual detectors operating during this period were within baseline values for both f090 and f150. The effective per-detector NET is skewed higher in the f090 band if the Mv9 wafer is included since this wafer had a low optical efficiency and therefore its detectors had higher NETs. Without Mv9 in the calculation, the pre-retrofit SAT1 focal plane had a median effective per-detector NET of $241\,\si{\micro\kelvin}\mathrm{_{CMB}\sqrt{s}}$; within 2\% of the goal value for the f090 band. This validated the overall optical performance of the instrument and that the detectors known to be good were operating well within expectations. In the f150 band, the effect of including Mv9 is small, shifting the median per-detector NET from 316 to $321\,\si{\micro\kelvin}\mathrm{_{CMB}\sqrt{s}}$. After the retrofit, the median per-detector NET for the f150 band is $336\,\si{\micro\kelvin}\mathrm{_{CMB}\sqrt{s}}$. This small increase is associated with the higher saturation powers in the f150 detectors leading to increased NEPs. Data taken from the good pre-retrofit detectors are planned to be included in all future cosmological analyses. 

The per-telescope NET comparison in Figure~\ref{fig:net_versus_model} shows the significant sensitivity improvement in the f090 band achieved by replacing the four detector wafers, bringing SAT1 from well above baseline to within baseline performance at 1\,mm PWV. The performance of the f150 band did not substantially change because the slight increase in the per-detector NETs was canceled out by the increase in overall detector yield.  

\bibliography{sources,sources_inflation}{}

\bibliographystyle{aasjournalv7}



\end{document}